\documentstyle[preprint,tighten,aps,psfig]{revtex}

\newcommand{\esla}{e\hspace{-0.45em}/}
\newcommand{\xisla}{\xi\hspace{-0.45em}/}
\newcommand{\etasla}{\eta\hspace{-0.45em}/}
\def\order#1{{\cal O}\left( #1 \right)}

\newcommand{\be}{\begin{equation}}
\newcommand{\ee}{\end{equation}}
\newcommand{\ba}{\begin{eqnarray}}
\newcommand{\ea}{\end{eqnarray}}

\newcommand{\non}{\nonumber}

\def\vec#1{{\mbox{\boldmath$#1$}}}

\newcommand{\p}{\mbox{$\vec{p}$}}

\newcommand{\k}{\mbox{$\vec{k}$}}

\newcommand{\r}{\mbox{$\vec{r}$}}
\newcommand{\e}{\mbox{$\vec{e}$}}

\newcommand{\lb}{\left (}
\newcommand{\rb}{\right )}

\newcommand{\ep}{\epsilon}
\newcommand{\vep}{\epsilon}

\newcommand{\om}{\omega}

\newcommand{\pop}[1]{\mbox{$\Lambda_+(#1)$}}
\newcommand{\nep}[1]{\mbox{$\Lambda_-(#1)$}}

\begin{document}

\preprint{BNL-HET-99/31, SLAC-PUB-8282, hep-ph/9910488}

\title{$\alpha^2$ corrections to
parapositronium decay: a detailed description}

\author{Andrzej Czarnecki\thanks{
e-mail:  czar@bnl.gov}}
\address{Physics Department, Brookhaven National Laboratory,\\
Upton, NY 11973}
\author{Kirill Melnikov\thanks{
e-mail:  melnikov@slac.stanford.edu}}
\address{Theoretical Physics Group, SLAC,\\
Mail Stop 81, P.O. Box 4349,
Stanford, CA 94309}

\author{ Alexander Yelkhovsky\thanks{
e-mail:  yelkhovsky@inp.nsk.su }
}
\address{ Budker Institute for Nuclear Physics,
\\
and Physics Department, Novosibirsk University,
\\
Novosibirsk, 630090, Russia}
\maketitle

\begin{abstract}
We present details of our recent calculation of $\alpha^2$ corrections
to the parapositronium decay into two photons.  These corrections are
rather small and our final result for the parapositronium lifetime
agrees well with the most recent measurement.  Implications for
orthopositronium decays are briefly discussed.
\end{abstract}

\pacs{36.10.Dr, 06.20.Jr, 12.20.Ds, 31.30.Jv}

\section{Introduction}

Precision measurements with nonrelativistic bound states provide a
sensitive test of bound state theory based on Quantum Electrodynamics
(QED).  Theoretical predictions generally agree well with experimental
results.  However, there are a few observables where the agreement is
not satisfactory.  The best known example is the lifetime of the
orthopositronium (o-Ps), where some experimental results
\cite{Westbrook89,Nico:1990gi}
differ from
the theory by several standard deviations.  The theoretical prediction
is not yet complete, because two-loop QED corrections to the o-Ps
decay have not been evaluated yet.  However, those corrections would
have to be very large to reconcile the theory with experiment.

Recently, we have reported a calculation of
corrections to the parapositronium (p-Ps) decay into two photons
\cite{Czarnecki:1999gv} in the second order in the fine structure
constant $\alpha$.  We argued there that the
smallness of the ${\cal O}(\alpha^2)$ corrections to the p-Ps lifetime
makes it unlikely that the analogous effects in o-Ps could alone
explain the discrepancy between theory and experiment.

The purpose of the present paper is to present a detailed description
of the calculation reported in \cite{Czarnecki:1999gv}.  We begin with
a brief discussion of the current theoretical and experimental status
of positronium (Ps) decays.

Theoretical predictions for p-Ps and o-Ps decay rates into 2 and 3
photons, respectively, can be expressed as series in $\alpha$:
\begin{eqnarray}
\Gamma_{\rm p-Ps}^{\rm theory} &=&
\Gamma^{(0)}_{p}
\left[ 1-\left(5-{\pi^2\over 4}\right) {\alpha\over \pi}
+2\alpha^2 \ln {1\over \alpha}
+ B_p \left( {\alpha\over\pi}\right)^2
- {3\alpha^3 \over 2\pi} \ln^2 {1\over \alpha}
+\ldots\right]
\label{eq:pps}
\\
\Gamma_{\rm o-Ps}^{\rm theory} &=&
\Gamma^{(0)}_{o}
\left[ 1-10.286\,606(10){\alpha\over \pi}
-{\alpha^2\over 3}\ln {1\over \alpha}
+ B_o \left( {\alpha\over\pi}\right)^2
- {3\alpha^3 \over 2\pi} \ln^2 {1\over \alpha}
+\ldots\right]
\label{eq:ops}
\end{eqnarray}
where
\begin{eqnarray}
\Gamma^{(0)}_{p}={m\alpha^5\over 2},\qquad
\Gamma^{(0)}_{o}={2(\pi^2-9)m\alpha^6\over 9\pi}
\label{eq:born}
\end{eqnarray}
are the lowest order decay widths of the p-Ps and o-Ps, respectively,
and the ellipses denote unknown
higher order terms which we will neglect in our analysis.  Corrections
of ${\cal O}(\alpha)$ were calculated in \cite{Harris57} for p-Ps.
For o-Ps the most accurate result was obtained in \cite{Adkins96},
where references to earlier works can be found.  The logarithmic
two-loop correction was found in \cite{Caswell:1979vz} for o-Ps and in
\cite{Khriplovich:1990eh} for p-Ps.  The leading logarithmic
correction at three loops was computed in \cite{DL}.  Some partial
results on the ${\cal O}(\alpha ^2)$ corrections for both p-Ps and
o-Ps can be found in
\cite{Adkins96,Burichenko:1992mk,Khriplovich:1994nq,%
Labelle:1994tq,Martynenko95:eng}.  A complete calculation of $B_p$ has been
reported recently in \cite{Czarnecki:1999gv} but the value of
$B_{o}$ has not been obtained so far.

Eqs.~(\ref{eq:pps},\ref{eq:ops},\ref{eq:born}) give
the following  predictions for the lifetimes:
\begin{eqnarray}
\Gamma_{\rm p-Ps}^{\rm theory} &=&
7989.42\,\mu {\rm s}^{-1}
  + \Gamma^{(0)}_p B_{p} \left( {\alpha\over\pi}\right)^2
 = \left(
 7989.42 +   0.043 \;B_p
   \right) \,\mu {\rm s}^{-1},
\label{eq:ptheor}
\\
\Gamma_{\rm o-Ps}^{\rm theory} &=& 7.0382\, \mu {\rm s}^{-1}
  + \Gamma^{(0)}_o B_{o} \left( {\alpha\over\pi}\right)^2
 =
\left(
 7.0382  +   0.39 \cdot 10^{-4} \; B_o
\right)
\,\mu {\rm s}^{-1}.
\label{eq:otheor}
\end{eqnarray}

On the experimental side, the Ann Arbor group
\cite{Westbrook89,Nico:1990gi} has found
\begin{eqnarray}
\Gamma_{\rm o-Ps}^{\rm exp} (\mbox{gas measurement}) &=& 7.0514(14)\,
\mu {\rm s}^{-1},
\nonumber \\
\Gamma_{\rm o-Ps}^{\rm exp} (\mbox{vacuum measurement}) &=& 7.0482(16)\,
\mu {\rm s}^{-1} ,
\label{eq:ann}
\end{eqnarray}
which, for $B_{o}=0$, differ from (\ref{eq:otheor}) by $9.4\sigma$ and
$6.3\sigma$ respectively.  This apparent disagreement of experiment
with theory has been known as the ``orthopositronium lifetime
puzzle'' for a long time. More recently, an independent
measurement by the Tokyo group found \cite{Asai:1995re}
\begin{eqnarray}
\Gamma_{\rm o-Ps}^{\rm exp} (\mbox{SiO$_2$ measurement}) &=& 7.0398(29)\,
\mu {\rm s}^{-1} ,
\label{eq:tok}
\end{eqnarray}
which agrees with the theory if $B_{o}$ is not too
large.\footnote{Another measurement of the o-Ps lifetime was
performed earlier in Mainz \cite{Hasbach:1987bh} with the result
$\Gamma_{\rm o-Ps}^{\rm exp} = 7.031(7)\, \mu {\rm s}^{-1}$.  This
result is consistent with the theoretical
predictions but uncertainties are clearly larger.  For a summary of
older experimental studies see \cite{AsaiPhD}.}
Thus, the present experimental situation is rather unclear.
If future experimental efforts confirm the Ann Arbor
results (\ref{eq:ann}), the orthopositronium
lifetime puzzle could be solved if $B_o$ turns out to be unusually
large, e.g.~$\sim 250$ for the vacuum measurement.  Alternatively, one
might speculate that some ``New Physics'' effects such as o-Ps decays
involving axions, millicharged particles, etc., cause the excess of
the measured decay rate over the QED predictions.  Some of those
exotic scenarios seem to have already been excluded by dedicated
experimental studies. (For a review and references to original papers
see e.g.~\cite{AsaiPhD}.)

p-Ps can decay into 2 photons and its lifetime is too short to be
measured directly.  For a long time its precise value remained
unknown.  However, it was realized in \cite{AlRam94} that such a
measurement would be very useful, since the calculation of the
coefficient $B_p$ is much easier than that of $B_o$. Therefore, p-Ps
offers an easier precision test of the bound-state QED.
This observation motivated the measurement of the p-Ps lifetime
\cite{AlRam94} which is $6.5$ times more accurate than
the best previous results.

To fully utilize that experimental result and enable the rigorous test
of bound-state QED envisioned in \cite{AlRam94}, we undertook a
complete calculation of the ${\cal O}(\alpha^2)$ corrections to p-Ps
rate \cite{Czarnecki:1999gv}.  We found that the non-logarithmic part
of those corrections is small; the coefficient $B_p$ is
\begin{eqnarray}
B_p= 5.1(3),
\label{eq:Bp}
\end{eqnarray}
and the theoretical prediction for the p-Ps lifetime becomes (half
of the logarithmic $\alpha^3$ term in (\ref{eq:pps})
is taken as an error estimate)
\begin{eqnarray}
\Gamma_{\rm p-Ps}^{\rm theory} &=& 7989.64(2)\,\mu {\rm s}^{-1}.
\end{eqnarray}
Comparing this number with the most  recent experimental result \cite{AlRam94}
\begin{eqnarray}
\Gamma_{\rm p-Ps}^{\rm exp} = 7990.9(1.7) \, \mu {\rm s}^{-1}
\label{eq:pexp}
\end{eqnarray}
we find excellent agreement between theory and experiment.

In the next Section we explain our approach to Ps decays with the
example of the leading order calculation.  In Sections \ref{sec3} and
\ref{sec4} we discuss, respectively, the so-called soft and hard
$\order{\alpha^2}$ contributions. The final result is
presented in Section \ref{secFinal}.

\section{Framework of the calculation}

Positronium is a bound state of an electron and a positron. Its energy
levels and lifetimes can be well understood within the framework of
nonrelativistic expansion in QED.
The decay width of p-Ps into two photons can be written as:
\be
\Gamma_{\rm p-Ps}
= \frac {1}{2!4\pi^2}\sum_\lambda \int \frac {{\rm d}^3 k_1}{2\om _1}
\frac {{\rm d}^3 k_2}{2\om _2} \delta (P-k_1-k_2)
  \left|
  \int \frac{{\rm d}^3p}{(2\pi)^3} {\rm Tr} [ A(\lambda, \p)~\Psi_P ]~\phi(\p)
  \right|^2,
\label{width}
\ee
where $P=(M_{\rm p-Ps},{\bf 0})$ is the four-momentum of the positronium,
$k_1$ and $k_2$  are photons' momenta,
$\phi(\p)$ is the positronium nonrelativistic
wave function, $A(\lambda,\p)$ is the amplitude
of the process $e^+ e^- \to 2 \gamma$,
$\lambda$ denotes polarizations of the photons, and
\be
\Psi_P = \frac {1+\gamma_0}{2\sqrt{2}}\gamma_5
\label{wf}
\ee
is the spinor part of the p-Ps wave function. The wave function
of the p-Ps is normalized to unity,
\be
\int { {\rm d}^3 p \over (2\pi)^3 } \phi(\p)^2 = 1,
\ee
hence the usual normalization factor, $1/2M_{\rm p-Ps}$, is absent in
Eq.~(\ref{width}).  From Eq.~(\ref{width}) one sees that $\Gamma_{\rm
p-Ps}$ is determined by two quantities: the annihilation amplitude of
the electron-positron pair into a pair of photons, and the positronium
bound state wave function. Since the typical velocity of electron and
positron in Ps is $\order{\alpha}$, the annihilation amplitude can be
expanded both in $\alpha$ and in the relative momentum $\p$ of the
electron and positron.  Corrections to the positronium wave function
are computed using the standard time-independent perturbation theory
and Breit correction to the nonrelativistic Hamiltonian.

We begin with the calculation of the decay rate to leading order.
Since the electron and positron in Ps are nonrelativistic, we employ
non-covariant perturbation theory.  Although it is not the only way,
the non-covariant technique permits an easier evaluation of the soft
corrections to the p-Ps decay rate.

The on-shell amplitude of the process $e^+e^- \to \gamma \gamma$ reads
\ba
A &=& \frac{8\pi\alpha E_p}{E_p+m}
    %v^{\dagger}
    \nep{\p}(\vec{\alpha}\e_2)
    \frac{\Lambda_+(\p-\k)-
    \Lambda_-(\p-\k)}{E_{p-k}}
    (\vec{\alpha}\e_1)\pop{\p}%w
 + (\e_2 \leftrightarrow \e_1,
    \k \leftrightarrow -\k)
      \non \\
  &=& \frac{8 \pi \alpha E_p}{E_p+m}
    \frac{%v^{\dagger}
    \nep{\p}(\vec{\alpha}\e_2)
    \lb \vec{\alpha}(\p-\k) + \beta m \rb
    (\vec{\alpha}\e_1)\pop{\p}%w
    }{E_{p-k}^2}
 + (\e_2 \leftrightarrow \e_1,
    \k \leftrightarrow -\k).
\label{A}
\ea
Here $E_p=\sqrt{m^2+\p^2}$ is the energy of the electron (or positron),
%$w$ and $v$ are bispinors of, respectively,   electron and positron
%at rest,  
$\Lambda_{\pm}(\p)$  are the projectors on the mass shell,
\be
\Lambda_{\pm}(\p) = \frac{1}{2} \lb 1 \pm
                \frac{\vec{\alpha}\p + \beta m}{E_p}
                \rb,
\ee
and $\k$ is the three-momentum of the photon in the final state.

To the leading order one can neglect the dependence
of the annihilation amplitude on the momentum $|\p| \sim m\alpha$,
which is small
in comparison with $m$ and $|\k| \sim m$. One obtains the following leading
order amplitude:
\be
A_{\rm LO} = -\frac {4\pi\alpha}{m^2} %\chi^+ 
(\vec{\alpha}\e_2)(\vec{\alpha}\k)(\vec{\alpha}\e_1) %\phi
.
\label{ALO}
\ee
{}From Eqs.~(\ref{ALO},\ref{wf}) it follows that the photon polarization
vectors are perpendicular to each other, i.e. $\e_1\e_2=0$.

Computing the trace in Eq.~(\ref{width}) and integrating
over $\p$  one obtains
\be
\left| \int \frac{{\rm d}^3p}{(2\pi)^3} {\rm Tr} [ A_{\rm LO}
(\lambda, \p)~\Psi_P ]~\phi(\p) \right|^2 =
\frac {32 \pi^2 \alpha^2}{m^2}\left|\psi(0)\right|^2,
\ee
which gives the following leading order  decay width:
\be
\Gamma_{0} = \frac {4 \pi \alpha^2}{m^2} \left|\psi(0)\right|^2 =
\frac {m\alpha^5}{2}.
\label{LO}
\ee

The calculation of higher order corrections to positronium lifetime
is performed in the framework of the Nonrelativistic Quantum
Electrodynamics (NRQED) \cite{Caswell:1979vz} which we regularize
dimensionally \cite{Pineda:1997bj}.  In this framework, all
contributions are divided into soft %(nonrelativistic) 
and hard %(relativistic) 
corrections. The soft contributions come from the
momenta region of the order of $k \sim m\alpha$ and thus are sensitive
to the details of the bound state dynamics. On the other hand, the
hard corrections arise as contributions of the relativistic momenta $k
\sim m$; their effect can be described by adding $\delta(\r)$-like
terms to the nonrelativistic Hamiltonian.

Special care in the present calculation  is required because of a single
$\gamma_5$ matrix in the positronium wave function.  Since a consistent
treatment of  single $\gamma_5$ is known to be a problem in
dimensional regularization, below we describe how we have dealt
with it.

We use the following fact: if the p-Ps decays into two photons
with polarizations  $\e_1$ and $\e_2$, then the photons are polarized
so that $\e_1 \e_2=0$. Let  the first photon be polarized
along the $x$  and the second photon along the $y$ axis; the
three-momentum of the first photon  $\k$  being along the $z$-axis.
If we introduce two auxiliary vectors
$$
\xi_\mu = \frac {P_\mu}{m},~~~~\eta_\mu = \frac {k_\mu}{m},
$$
then the standard four-dimensional representation of  the
$\gamma_5$ matrix, $\gamma_5 = i\gamma_0 \gamma_x \gamma_y \gamma_z$,
can  be written as:
\be
\gamma_5 = \frac {i}{2m^2} \xisla  \esla_1 \esla_2 \etasla.
\label{g5}
\ee
We use this equation to define  $\gamma_5$ in $d$ dimensions.

With an additional trick it is possible to avoid any reference to
photon polarization vectors. Let us consider Eq.~(\ref{width}) and use there
the explicit representation for the $\gamma_5$ matrix, Eq.~(\ref{g5}).
Then the following trace has to be computed:
\be
{\rm Tr} [\xisla \esla_1 \esla_2 \etasla
 M_{\alpha \beta}]~e_1^\alpha e_2^\beta.
\label{3}
\ee
Taking into account that  $e_1e_2 = e_{1,2}k=e_{1,2}P=0$,
one concludes  that the final result can only depend on $e_1^2$ and $e_2^2$.
In this situation, one can average Eq.~(\ref{3}) over directions
of $e_1$ and $e_2$, provided one respects the above constraints.
We arrive  at the following formula:
\be
{\rm Tr} [\xisla \esla_1 \esla_2 \etasla
 M_{\alpha \beta} ]~e_1^\alpha e_2^\beta
\to
{\rm Tr} [\xisla   \gamma^\nu \gamma^\mu \etasla
 M_{\alpha \beta} ]~T_{\mu \nu}^{\alpha\beta},
\ee
where
\ba
&&T_{\mu\nu}^{\alpha\beta} =
\frac {
 (3-2\ep)d_\mu^ \alpha d_\nu^\beta
   -d_{\mu \nu}d^{\alpha \beta}-d_\mu^\beta d^\alpha_\nu}
{4(1-2\ep)(2-\ep)(1-\ep)},
\nonumber \\
&&d_{\mu \nu} \equiv g_{\mu \nu}-\xi_\mu \xi_\nu
 -\eta_\mu \eta_\nu.
\ea

According to  the discussion of  different momentum regions
contributing to the second order corrections, we divide the coefficient
$B_{p}$ in Eq.~(\ref{eq:ptheor}) into three parts:
\be
B_{p} = B_{p}^{\rm squared} + B_{p}^{\rm hard} + B_{p}^{\rm soft},
\ee
where $B_{p}^{\rm squared}$ is the contribution of the one-loop
amplitude squared and $B_{p}^{\rm {hard,~soft}}$ are the hard and soft
two-loop contributions.  The square of the one-loop amplitude
is easily obtained from the one-loop result:
\be
B_p^{\rm squared} = \left ( \frac {5}{2} - \frac {\pi^2}{8} \right )^2
 = 1.6035.
\label{oneloop2}
\ee

In contrast to the first order correction which
arises from hard photon exchange only,
the second order correction is more difficult to compute, because of
the appearance of the soft scale effects.  In addition,
hard and soft corrections are not finite separately.
Below we discuss the calculation of these corrections.

\section{Soft scale contributions}
\label{sec3}
As follows from Eq.~(\ref{width}), in order to obtain
$B_{p}^{\rm soft}$ one has to  compute
relativistic corrections to the annihilation amplitude
$e^+e^- \to \gamma \gamma $  and relativistic corrections
to the positronium wave function induced by the Breit Hamiltonian.
Accordingly,  the soft contribution is separated into two pieces:
\be
B_p^{\rm soft} = B_p^{\rm soft}(AA) +  B_p^{\rm soft}(WF).
\ee

\subsection{Relativistic corrections to the amplitude}

The calculation of  relativistic corrections to the amplitude is
straightforward. One starts with the on-shell amplitude Eq.~(\ref{A})
and expands it up to relative order ${\cal O}(\p^2/m^2)$.
The calculation of these corrections
can be performed in three dimensions. To demonstrate this,
let us write the correction  to the amplitude in the
form:\footnote{A linear term in the expansion of the amplitude in
$\p$ does not contribute to the decay rate
since the positronium ground state wave function is spherically
symmetric.}
\be
\delta A = A^{(2)}_{ij} p_i p_j.
\ee
To calculate the correction to the p-Ps lifetime induced by
$\delta A$, we have to compute the following integral:
\be
\int \frac {{\rm d}^dp}{(2\pi)^d} \phi(\p) \delta A
\equiv
\frac {1}{d} A^{(2)}_{ii}
\int \frac {{\rm d}^dp}{(2\pi)^d} \phi(\p) \p^2.
\ee
To this end we use the Schr\"odinger
equation in the momentum space:
\be
\int \frac {{\rm d}^dp}{(2\pi)^d} \phi(\p) \p^2 =
\int \frac {{\rm d}^dp}{(2\pi)^d}  \frac {4\pi \alpha m \p^2}{\p^2-mE}
\int \frac {{\rm d}^dk}{(2\pi)^d} \frac {\phi(\k)}{(\p - \k)^2}.
\ee
Rewriting
$$
 \frac {\p^2}{\p^2-mE} = 1+\frac {mE}{\p^2-mE},
$$
shifting the integration momenta in the first term $\p \to \p + \k$
and using the fact that the scaleless integrals vanish in dimensional
regularization, we arrive at
\be
\int \frac {{\rm d}^dp}{(2\pi)^d} \phi(\p) \p^2 = mE \psi(0).
\label{linear}
\ee

We see that the amplitude $A^{(2)}_{ij}$ is needed only
in the $\ep \to 0$ limit, where it can be easily calculated.
We obtain:
\be
\delta A = -\frac {2\p^2}{3m^2} A_{\rm LO},
\ee
which induces the following ${\cal O}(m\alpha^2)$
correction to the p-Ps lifetime:
\begin{eqnarray}
B_p^{\rm soft}(AA) = {\pi^2\over 3}.
\label{ga}
\end{eqnarray}

Recently there has been some discussion in the literature
\cite{Labelle:1994tq,Khriplovich:1994nq,Khriplovich:1996ru} concerning
the linearly divergent integral in Eq.~(\ref{linear}).  Our approach
to the linear divergence is based on dimensional regularization which
permits a consistent treatment of hard and soft corrections
simultaneously.  We dealt similarly with linearly divergent integrals
in our recent calculations of the ${\cal O}(m\alpha^6)$ corrections to
positronium $S$-wave energy spectrum
\cite{Czarnecki:1998zv,Czarnecki:1999mw} and found agreement with
earlier results obtained in a different regularization scheme
\cite{Ph}.  This gives us confidence in the result given in
Eq.~(\ref{linear}).

\subsection{Relativistic corrections to the wave function}

Relativistic corrections to the positronium wave function
can be computed using Breit Hamiltonian. Since we regularize
all divergences dimensionally, we need that Hamiltonian
in $d$ dimensions, as it has been derived
in \cite{Czarnecki:1999mw}.

Breit Hamiltonian  projected on the $S$-states reads
\be\label{HBr}
U(\r,\p) = - \frac{ \p^4 }{ 4m^3 }
              + \frac{ d-1 }{ 4m } \left\{ \frac{ \p^2 }{ m },C(r) \right\}
              + \frac{ d\pi\alpha }{ m^2 }\delta(\r)
              - \frac{ \pi\alpha }{ 4dm^2 }
                [\sigma_i,\sigma_j][\sigma'_i,\sigma'_j] \delta(\r).
\ee
Here the Pauli matrices $\sigma_i$ and $\sigma'_i$ act
on the two-component spinors of the nonrelativistic
electron and positron respectively, and
\be
C(r) = - \frac{ \alpha\Gamma(d/2-1)
            }{ \pi^{d/2-1} r^{d-2} }
\ee
is the $d$--dimensional generalization of the Coulomb potential.
The operator $U(\r,\p)$ induces the following correction
to the p-Ps width,
\be
\frac{\delta_{\rm B} \Gamma}{\Gamma_{\rm LO}}
    = \Delta_1 + \Delta_2,
\ee
where the first term is due to the $\delta(\r)$-part
of the operator $U$ and the second term is due to the remaining
terms of that operator. We note also,
that $\Gamma_{\rm LO}$ in the above formula
stands for the leading order decay width computed in $d$-dimensions,
in contrast to the three-dimensional result Eq.~(\ref{LO}).
All necessary formulas can be extracted from the calculation
described after Eq.~(36) of Ref. \cite{Czarnecki:1999mw}.
One obtains:
\ba
\Delta_1 &=& -\frac {\alpha^2}{8} \left ( \frac {1}{\ep} - 4
\ln (m\alpha) -2 \right ),
\nonumber \\
\Delta_2 &=& \frac {5\alpha^2}{8} \left ( \frac {1}{\ep}
- 4 \ln (m \alpha)  + \frac {31}{5} \right ).
\ea
Thus the wave function correction contribution to the positronium
lifetime becomes
\be
B_p^{\rm soft}(WF)
+2\pi^2\ln{1\over \alpha}
= {\pi^2\over 2\epsilon} + 2 \pi^2 \ln{1 \over m\alpha} + \frac
{33\pi^2}{8} ,
\label{gb}
\ee
On the LHS of the above equation  we have separated the logarithm
of the fine structure constant  to be consistent with division of corrections
introduced in Eq.~(\ref{eq:pps}).

\subsection{Final result for the soft contributions}

The sum of the corrections to the
annihilation amplitude (\ref{ga})
and to the wave function (\ref{gb}) gives the final result for the
soft contributions,
\begin{equation}
B_p^{\rm soft} =
 \frac {\pi^2}{2\epsilon} - 2\pi^2 \ln m
+ \frac{107\pi^2}{24}.
\label{soft}
\end{equation}

\section{Hard scale contribution}
\label{sec4}
\begin{figure}[h]
\hspace*{-7mm}
\begin{minipage}{16.cm}
\[
\mbox{
\begin{tabular}{cccc}
\psfig{figure=  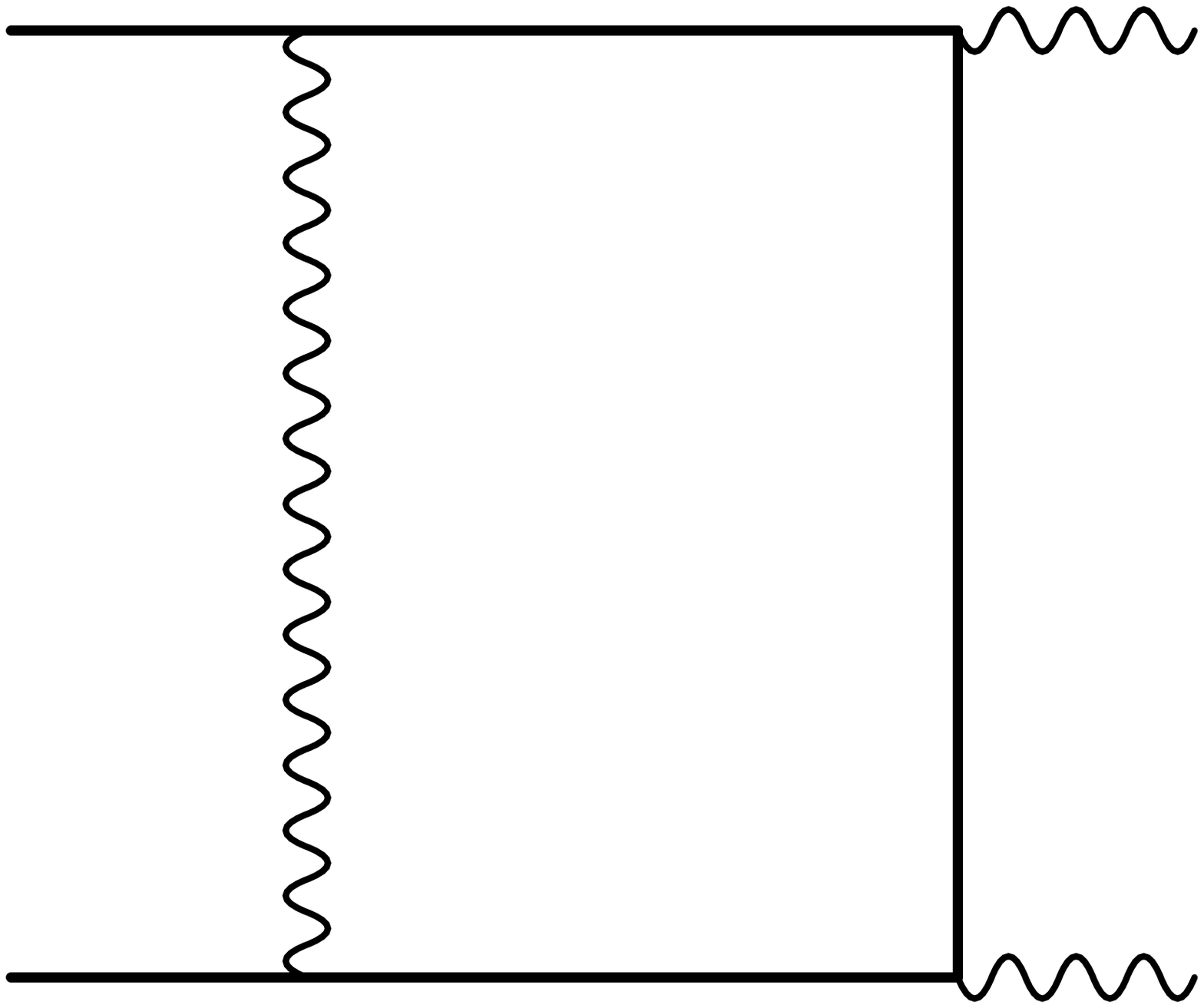,width=25mm}
&\hspace*{5mm}
\psfig{figure=  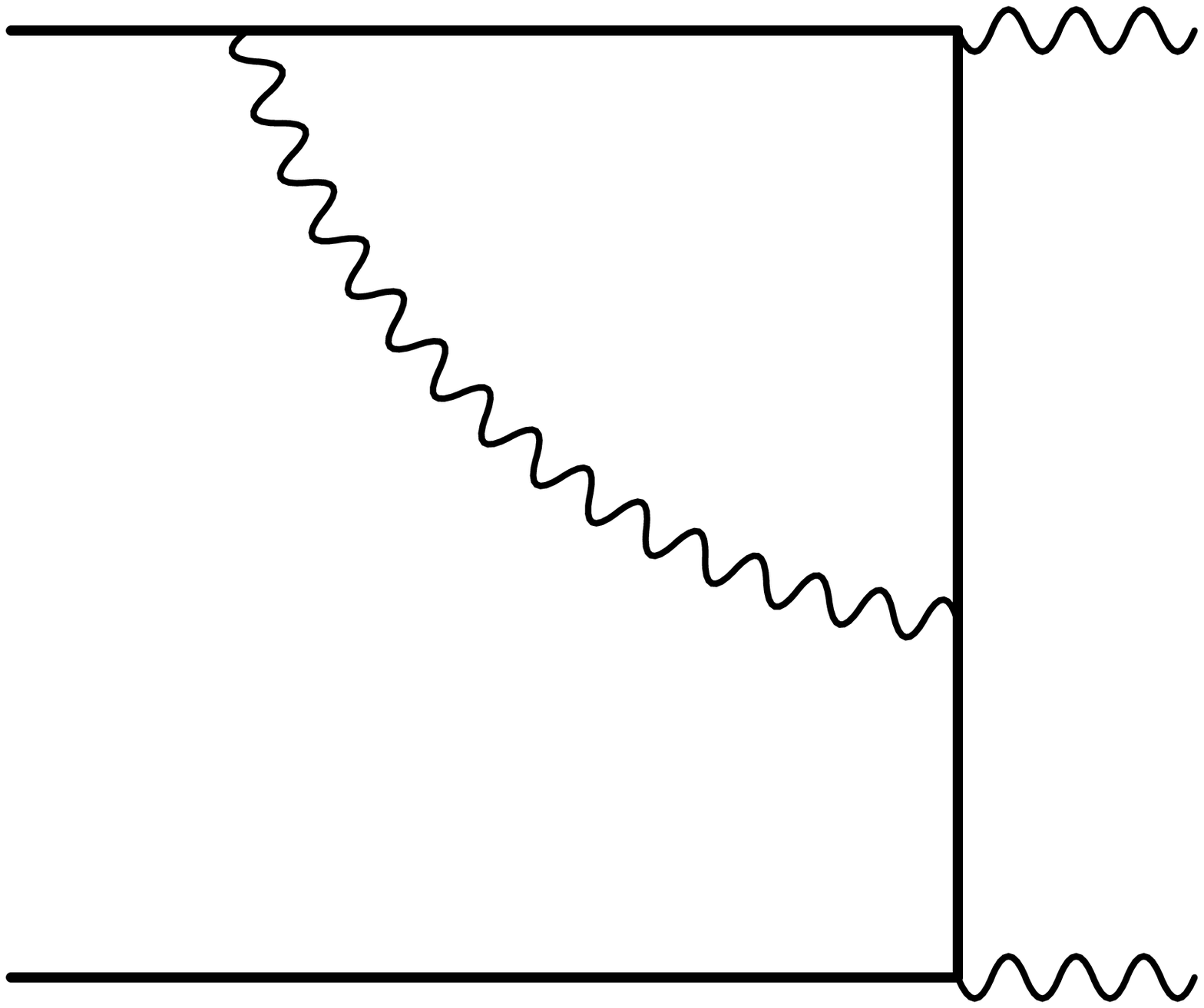,width=25mm}
&\hspace*{5mm}
\psfig{figure=  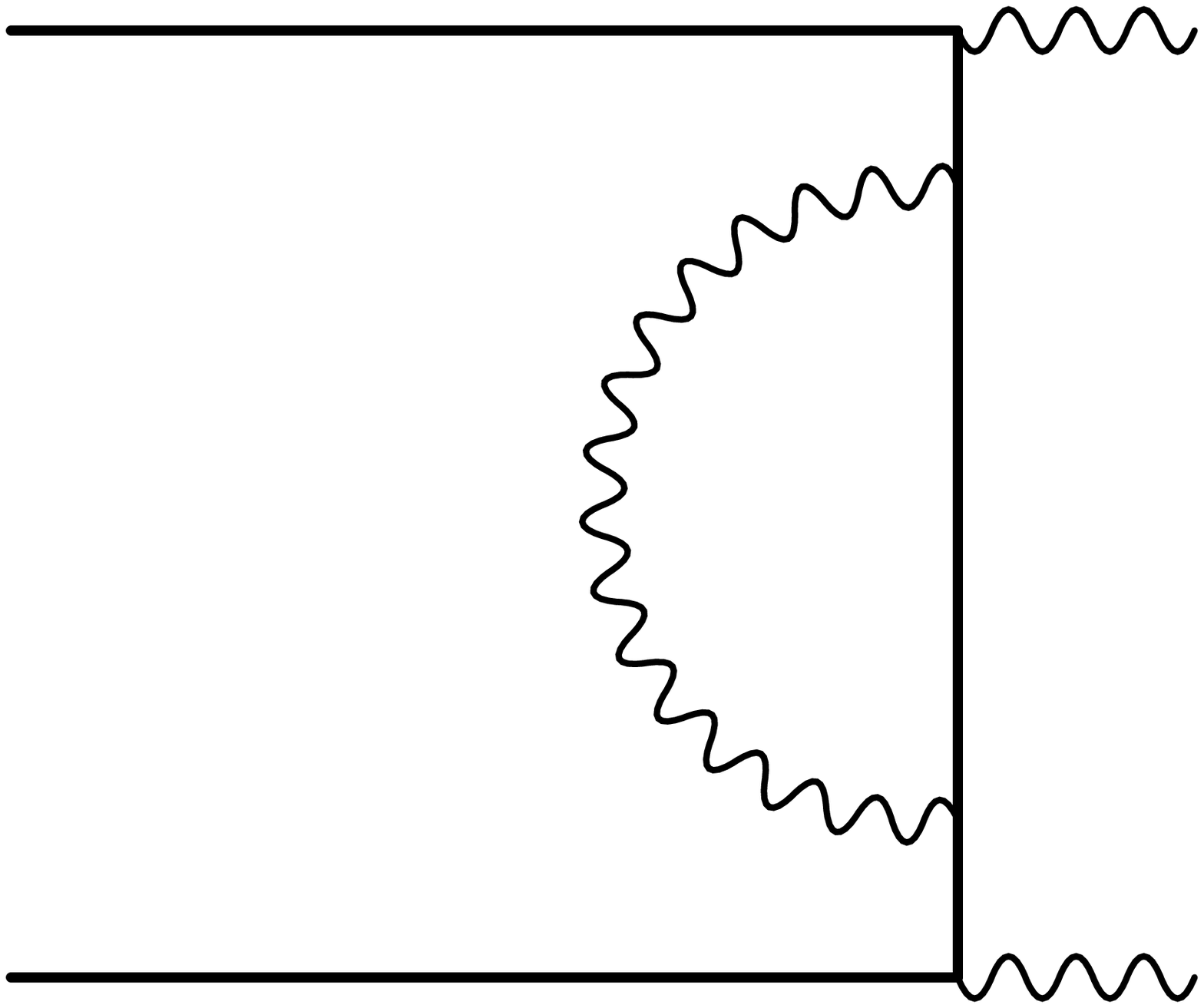,width=25mm}
&\hspace*{5mm}
\psfig{figure=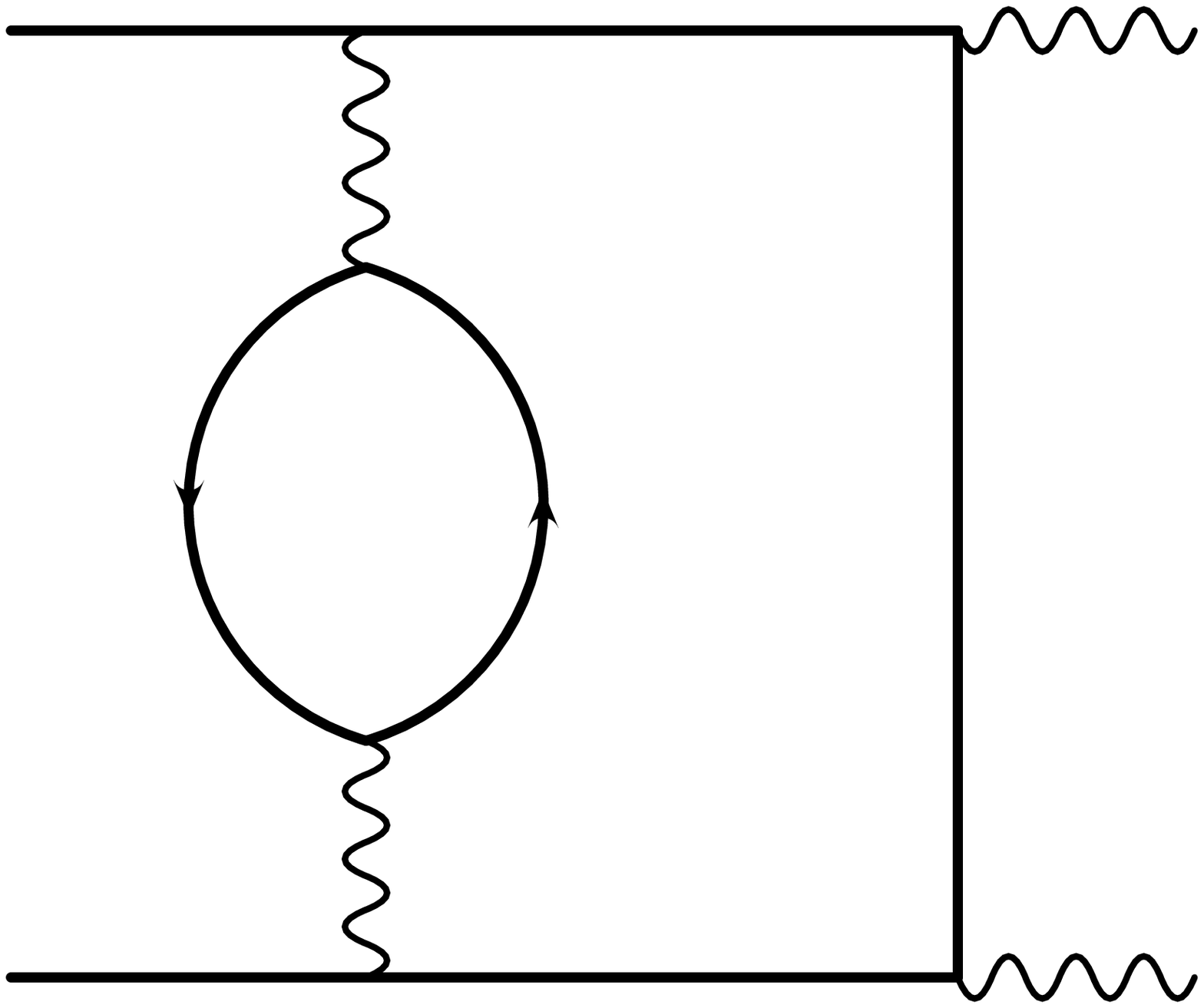,width=25mm}
\\[1mm]
 $S_1$ &  $S_2$ &  $S_3$ & VP
\\[1mm]
\end{tabular}
}
\]
\end{minipage}
%\vspace*{10mm}
\caption{One-loop corrections to p-Ps decay.}
\label{fig:onePs}
\end{figure}

The second class of corrections are the hard scale contributions.
The corresponding Feynman diagrams are shown in
Figs.~\ref{fig:onePs}(VP), \ref{fig:light},
\ref{fig:twoPs}, \ref{fig:twoCT}.
They should be computed in dimensional regularization,
with external electron and positron at rest.

$B_p^{\rm hard}$ consists of three types of contributions:
vacuum polarization insertions in the photon propagators,
light-by-light scattering diagrams, and two-photon corrections to the
annihilation amplitude,
\begin{eqnarray}
B_p^{\rm hard}=B_p^{\rm hard}(\mbox{VP})
+ B_p^{\rm hard}(\mbox{LL})
+ B_p^{\rm hard}(\gamma\gamma).
\end{eqnarray}

Vacuum polarization insertions into the one-loop
graphs (an example is shown in Fig.~\ref{fig:onePs}(VP))
were computed in \cite{Burichenko:1995as,AdkShif},
\begin{equation}
B_p^{\rm hard}(\mbox{VP})= 0.4473430(6).
\label{VP}
\end{equation}

\subsection{Light-by-light scattering contributions}

\begin{figure}
\begin{minipage}{16.cm}
\[
\mbox{
\begin{tabular}{cc}
\psfig{figure= 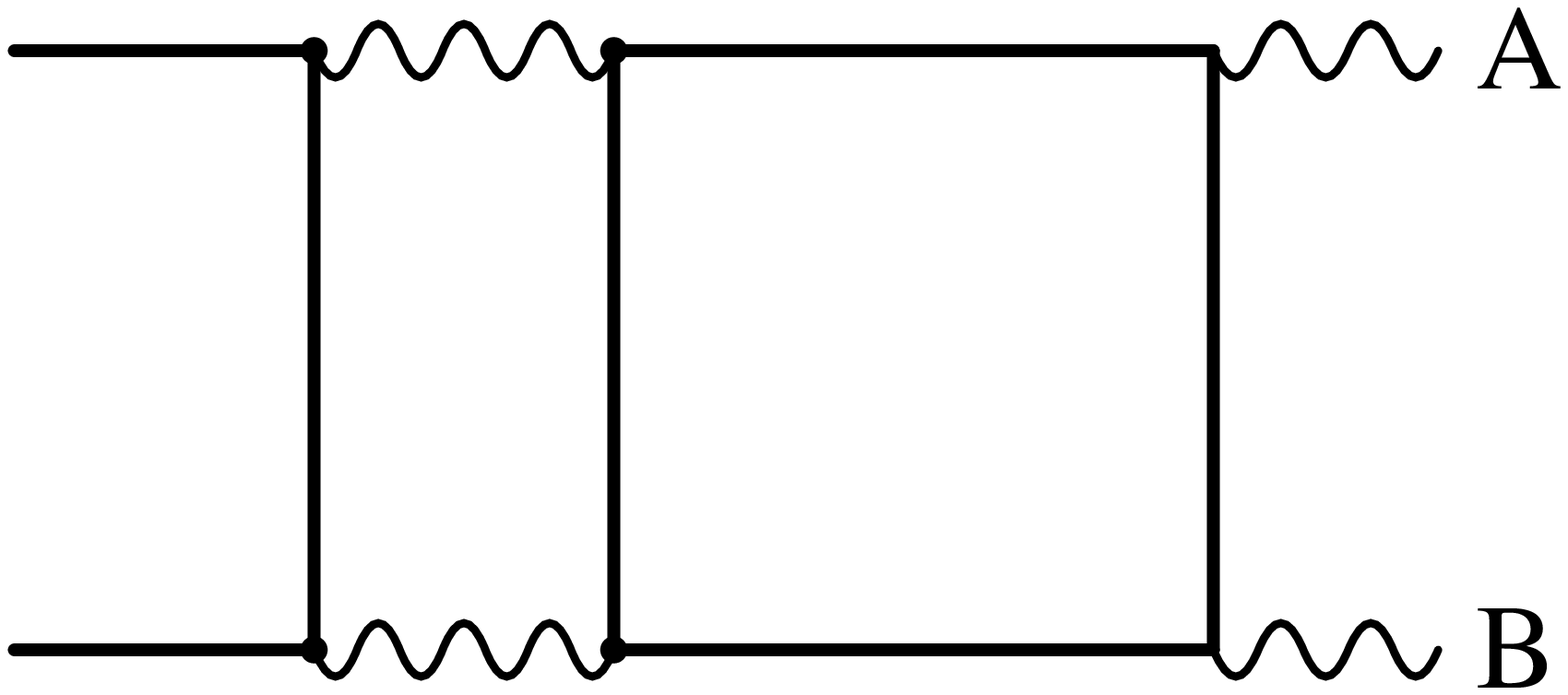,width=35mm}
&\hspace*{5mm}
\psfig{figure= 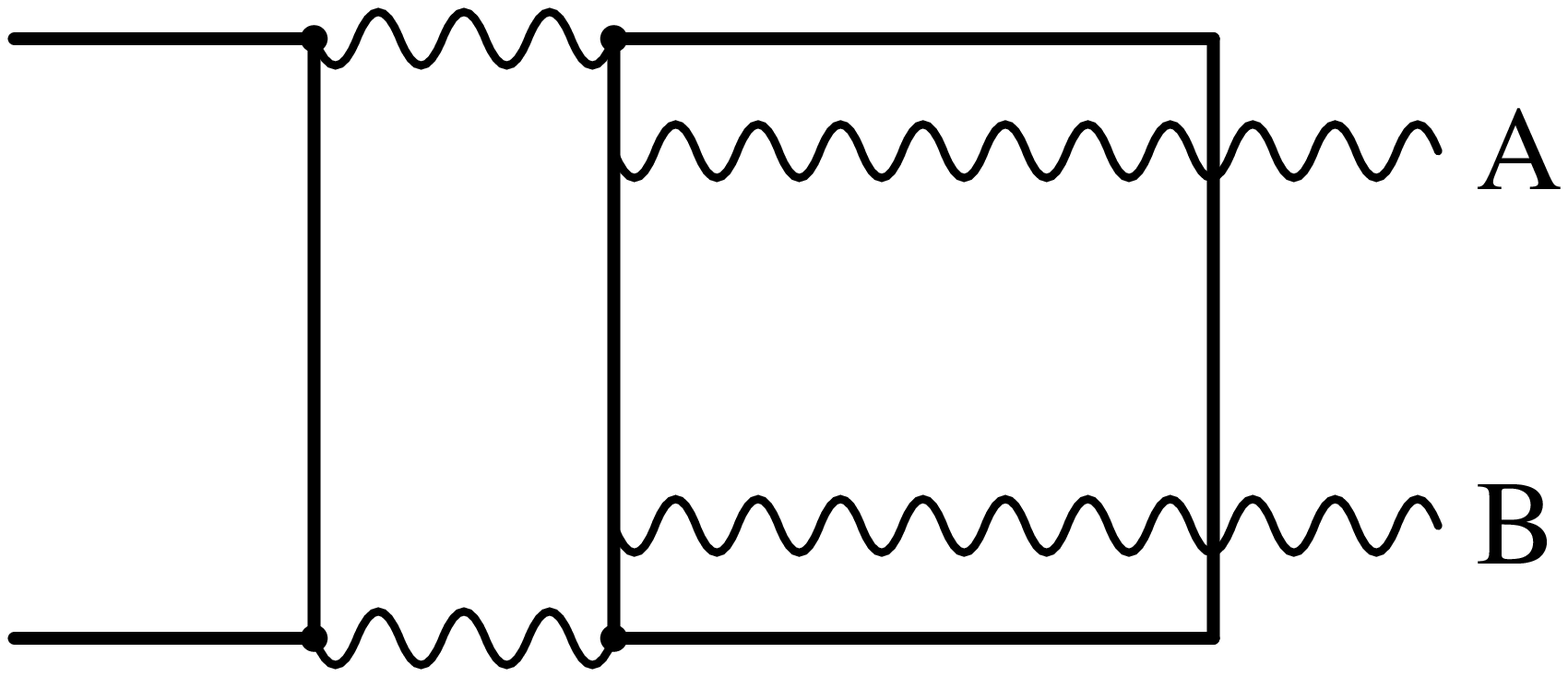,width=35mm}
\\
(a) & (b) \\[4mm]
\psfig{figure= 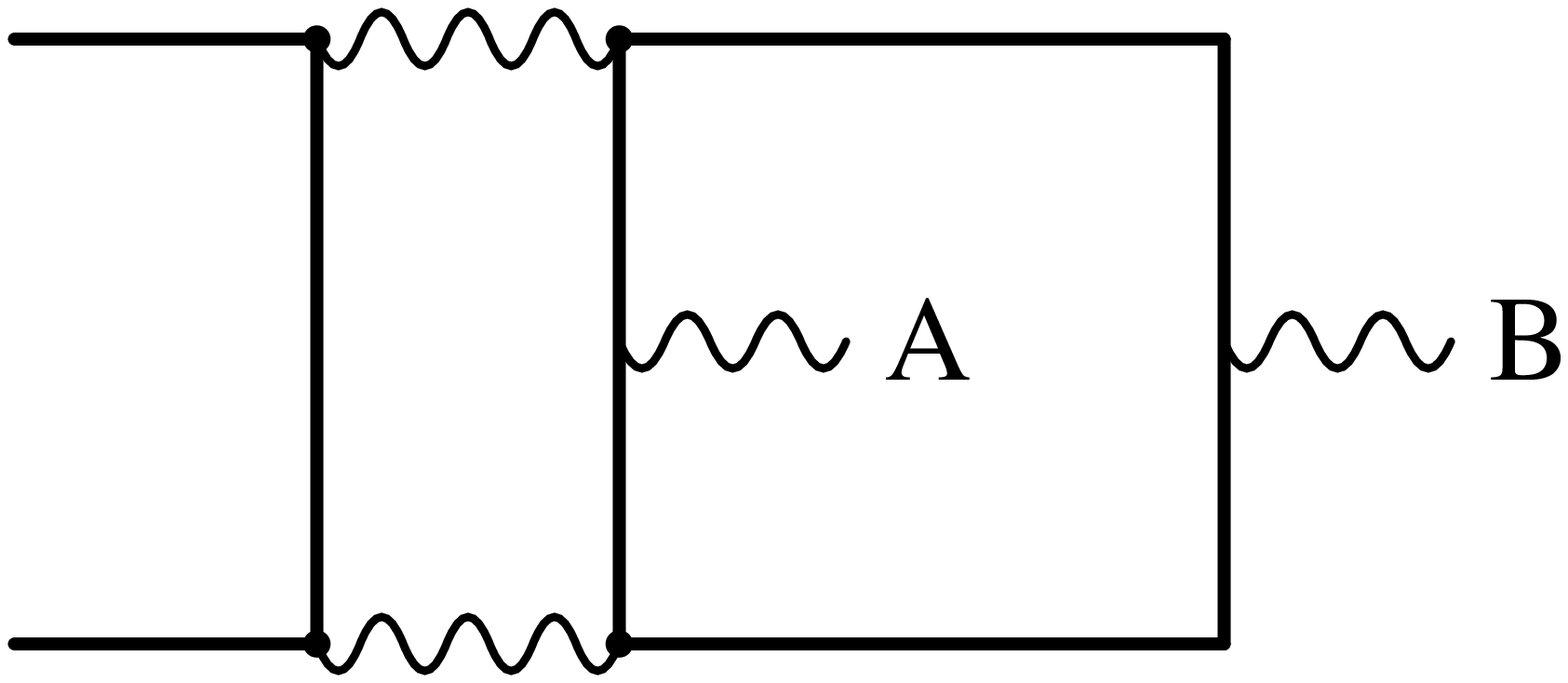,width=35mm}
&\hspace*{5mm}
\psfig{figure= 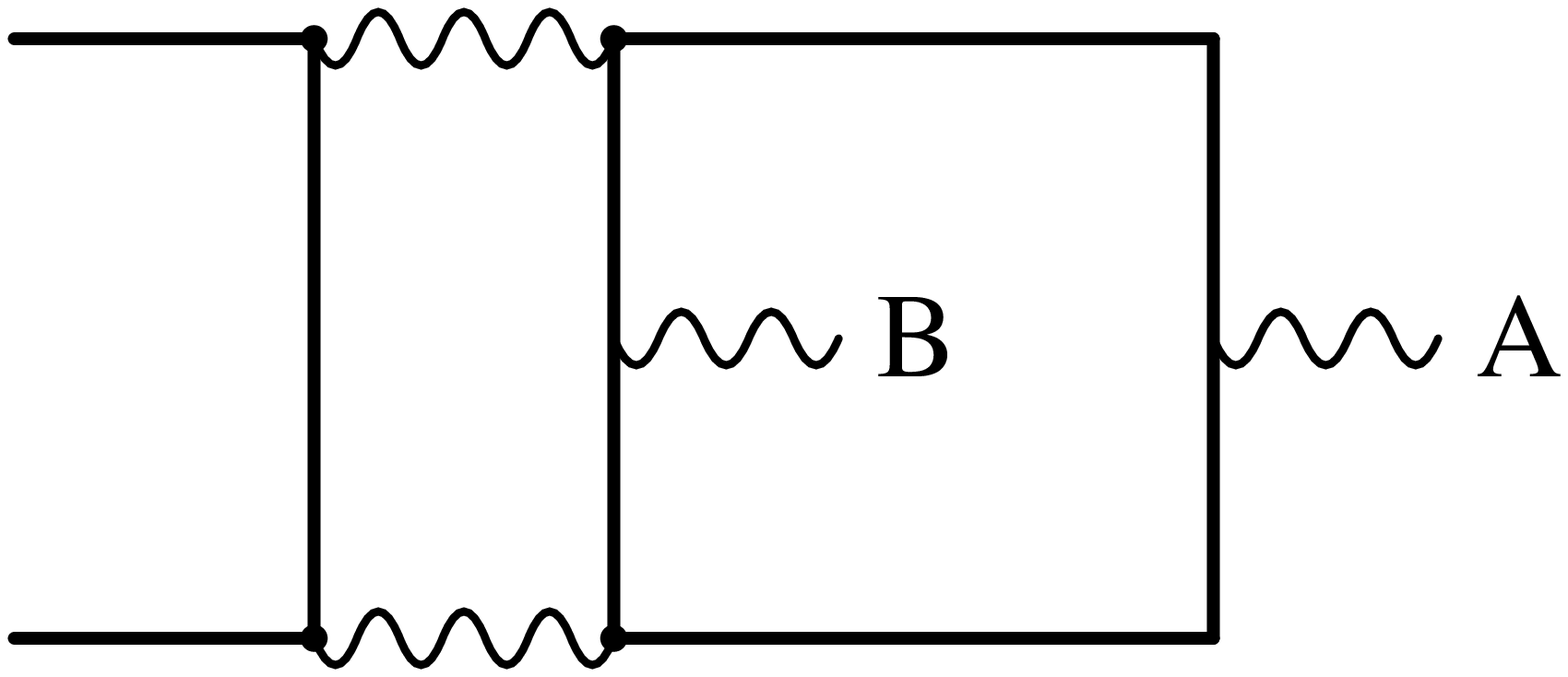,width=35mm}
\\
(c) & (d) \\[2mm]
\end{tabular}
}
\]
\end{minipage}
\caption{Light-by-light scattering contributions to p-Ps decay.}
\label{fig:light}
\end{figure}

One class of the second order corrections to the p-Ps lifetime arises
from the photon-photon scattering, shown in Fig.~\ref{fig:light}.
These contributions are relatively small; it is interesting, however,
that the positronium lifetime measurement of increased precision may
become sensitive to effects of non-linear QED.

For the diagrams shown in Fig.~\ref{fig:light}, we find that the
planar ones are equal (a=b), and so are the non-planar ones (c=d).
Further, both classes remain unchanged when we cross the internal
photons.  Therefore we only need to compute two diagrams, one planar
and one non-planar.  We will call them $L_1$ and $L_2$.  The
contribution of those diagrams to the final expression for the
amplitude is $4L_1 + 2 L_2$.  The factor 4 arises because we have two
types (a) and (b) which differ by the orientation of the fermion loop,
and both can have parallel or crossed internal photon lines.  For the
type $L_2$ we only have a factor 2 for the orientation of the fermion
loop, because crossing of the internal photon lines simply exchanges
(c) and (d).

The light-by-light diagrams in Fig.~\ref{fig:light}
have an imaginary part because of the two-photon cut.  This complicates
a numerical integration over Feynman parameters and  we
used a different method to evaluate them. The idea is to (formally)
assign a large mass $M$ to the internal fermion line. The diagram
may then be expanded in ratio $m/M$ using the so-called
Large Mass Expansion
\cite{Chetyrkin91,Smirnov:1995tg,Tkachev:1994gz}. This reduces the task
to the calculation of two-loop vacuum or massless propagator
integrals.  The price to be paid is that the result
is a series in $m/M$, while we are interested in the
value of the series at $m/M=1$.

Fortunately, using symbolic manipulation programs one can compute many
terms of the series; in our calculation about twenty terms were
computed for each diagram. The resulting series converge well,
especially for $L_2$, and the number of computed terms is sufficient
to obtain an accurate estimate of this contribution at
$x=1$.  The behavior of the series for $L_1$ is improved if we make a
change of variables $m/M = \sqrt{z/(2-z)}$.  The $n-$th term of the
series in variable $z$ decreases roughly like  $\ln^2 n/n^2$.  
Finally, we find
\ba
L_1 &=& 0.494(30),
\qquad L_2 = -0.348(20),
\label{l12}
\ea
and the contribution of these diagrams to the coefficient $B_p$ is:
\begin{equation}
B_p^{\rm hard}(\mbox{LL}) = 4L_1+2L_2  =  1.28(13).
\label{lbl}
\end{equation}

\subsection{Two-photon corrections}

\begin{figure}[h]
\hspace*{-2mm}
\begin{minipage}{16.cm}
\[
\mbox{
\begin{tabular}{ccccc}
\psfig{figure=  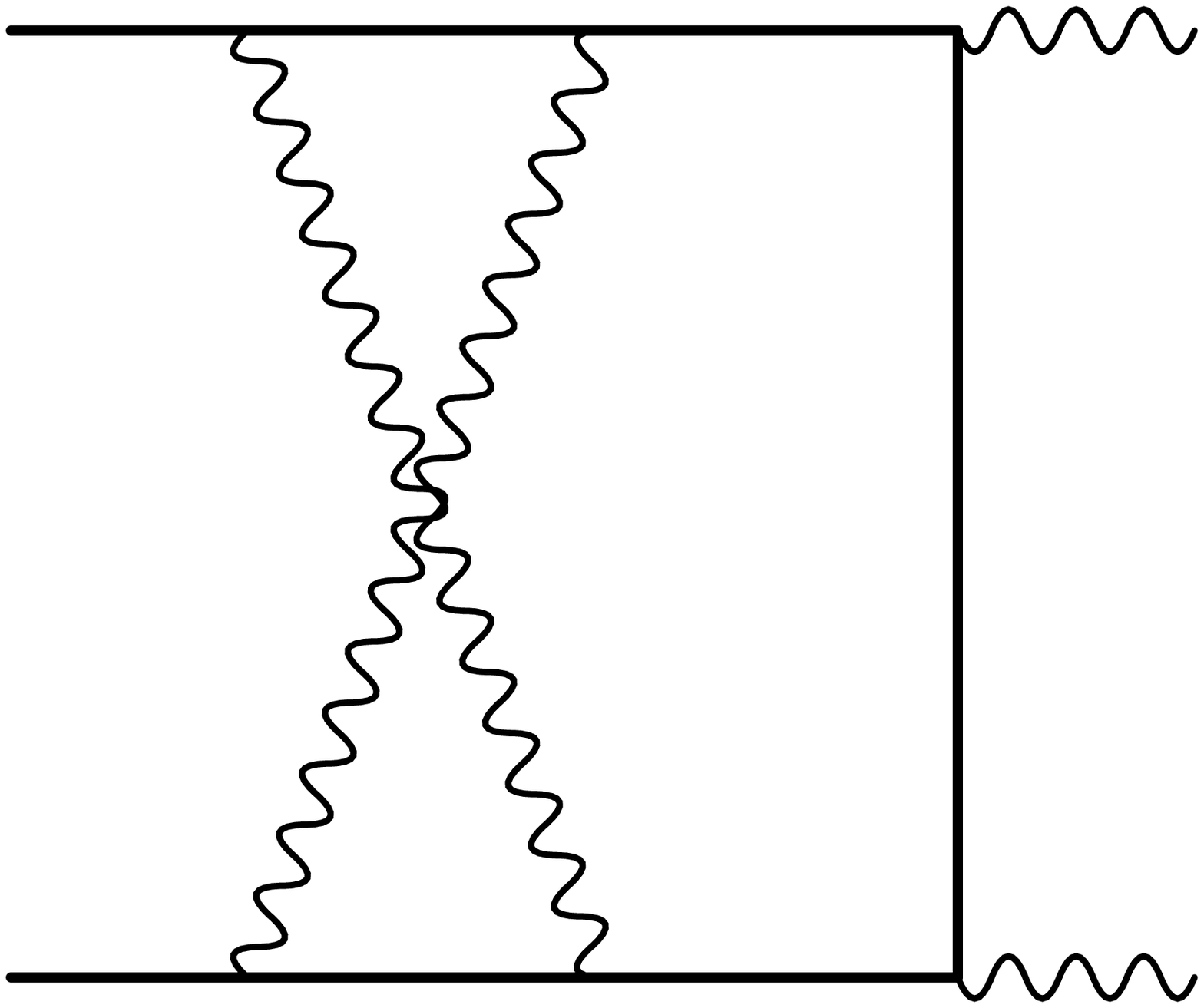,width=25mm}
&\hspace*{2mm}
\psfig{figure=  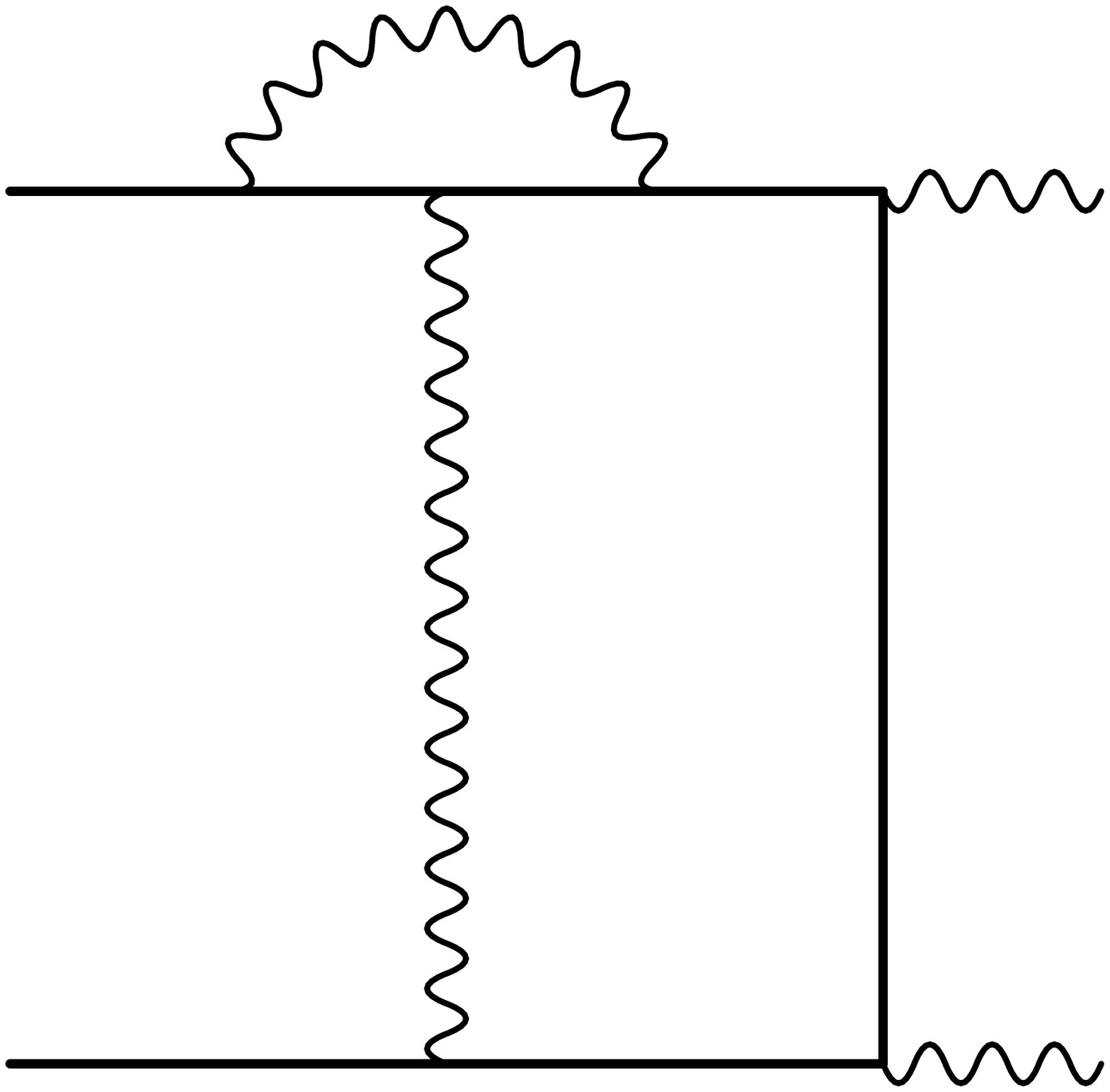,width=25mm}
&\hspace*{2mm}
\psfig{figure=  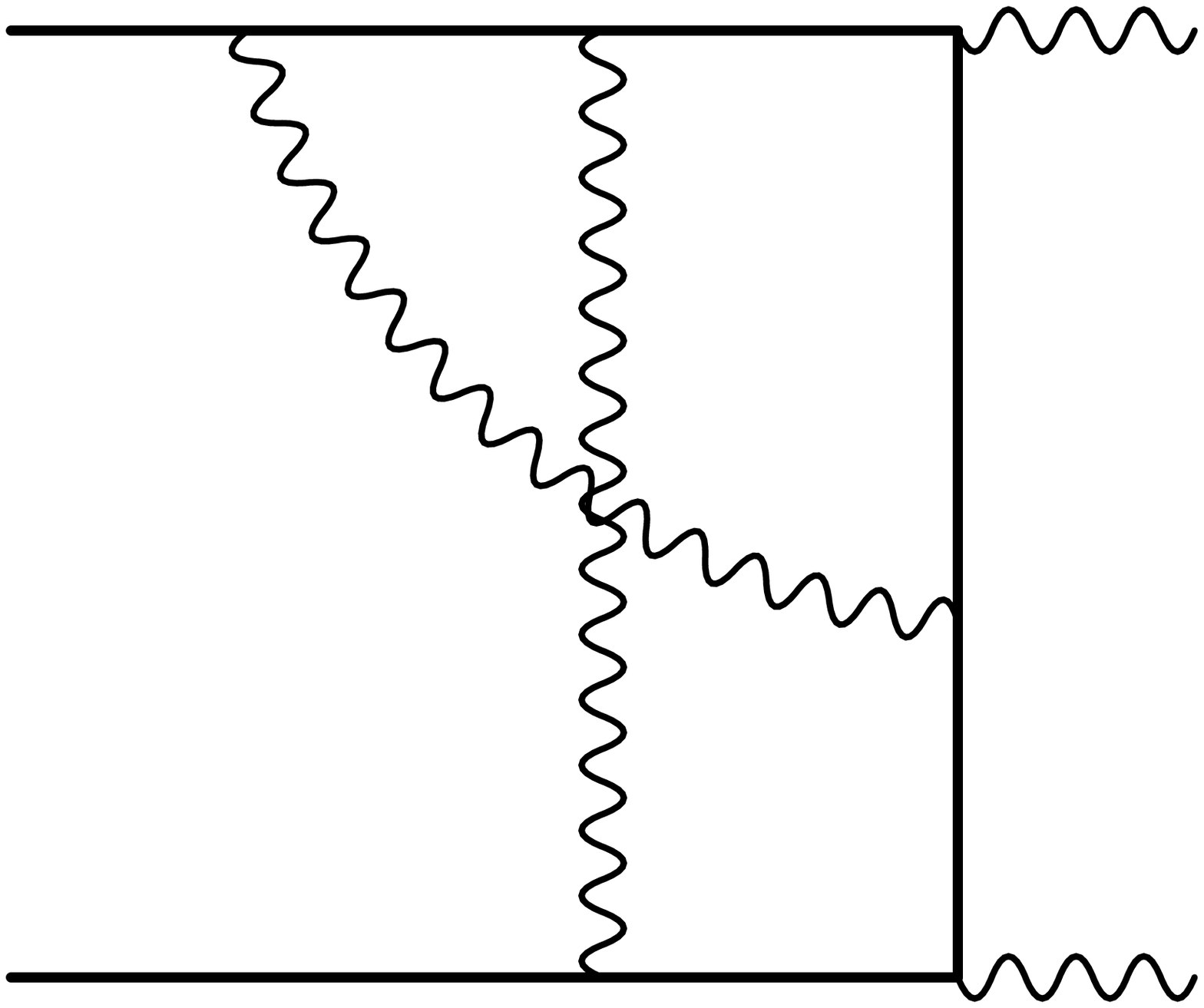,width=25mm}
&\hspace*{2mm}
\psfig{figure=  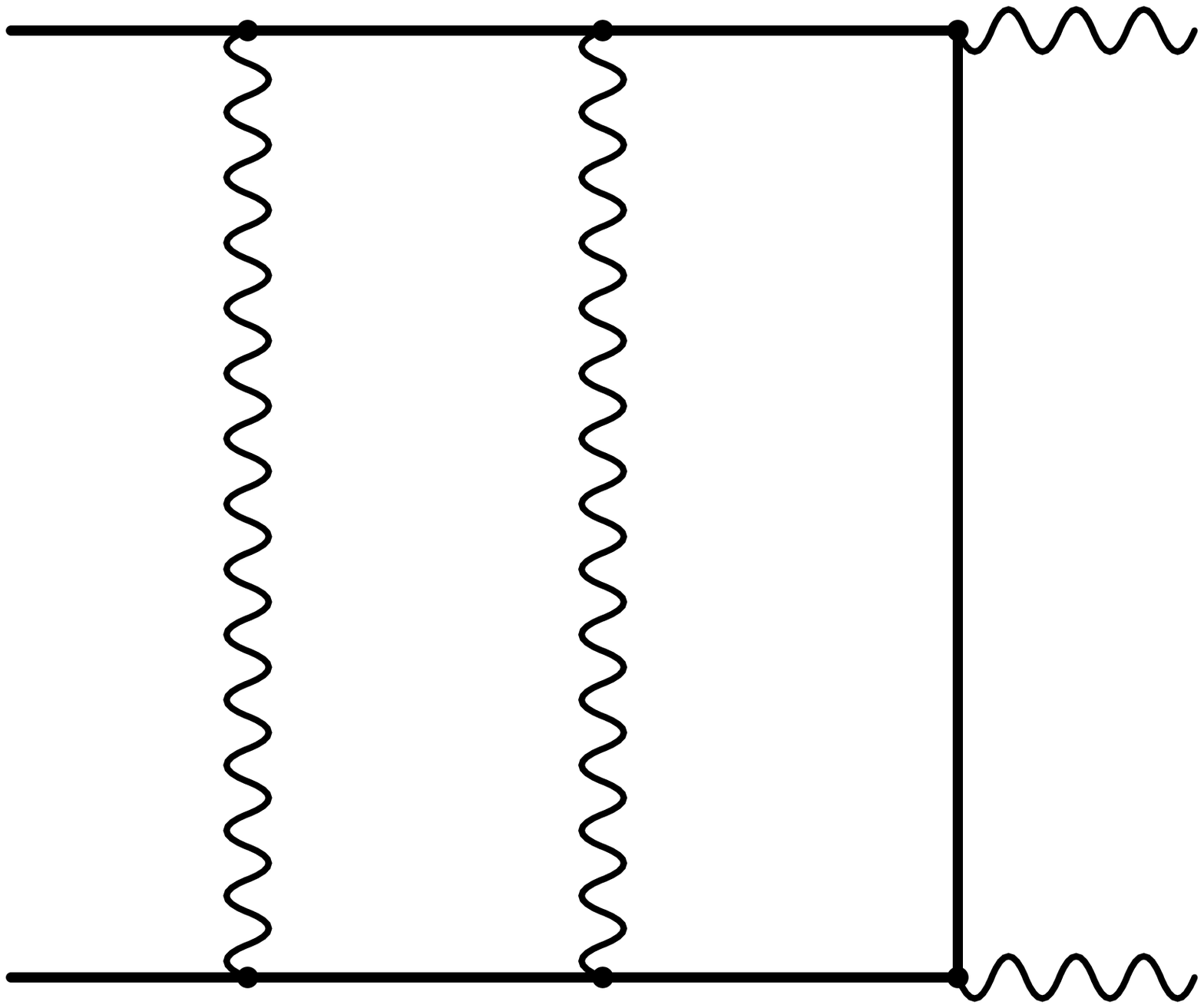,width=25mm}
&\hspace*{2mm}
\psfig{figure=  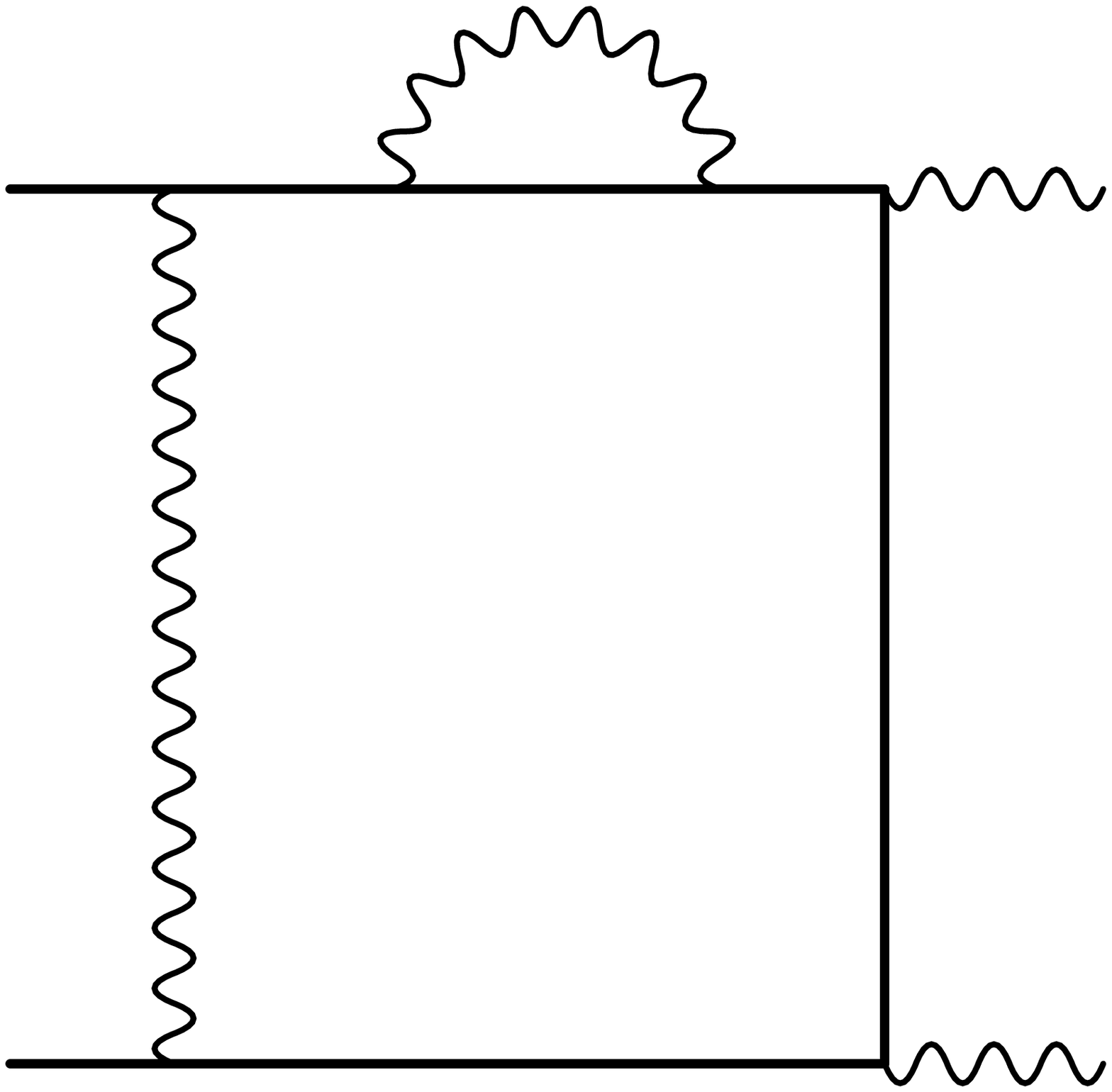,width=25mm}
\\[1mm]
 $D_1$ & $D_2$ & $D_3 $ & $D_4$ & $D_5$
\\[1mm]
%%%
\psfig{figure=  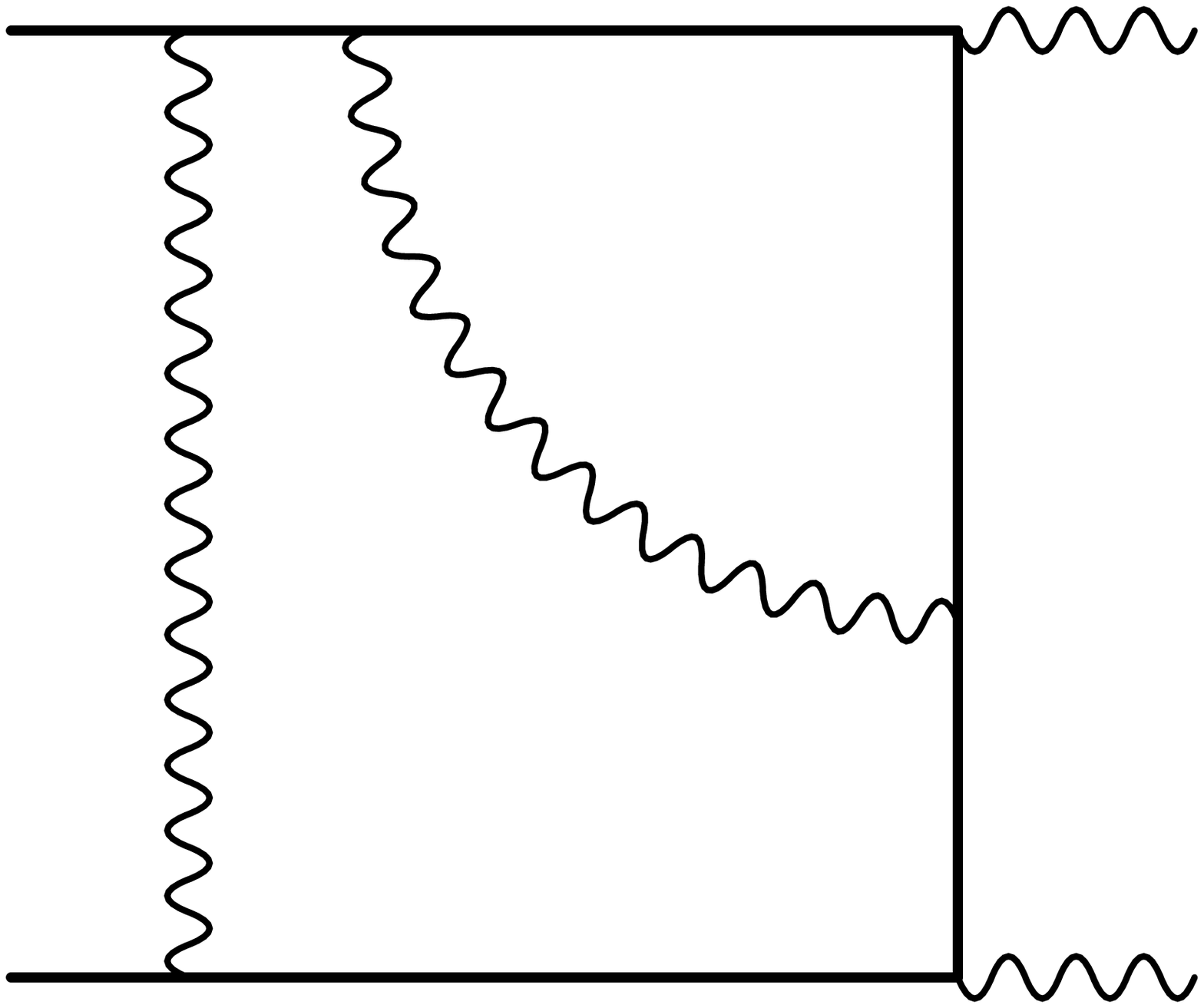,width=25mm}
&\hspace*{2mm}
\psfig{figure=  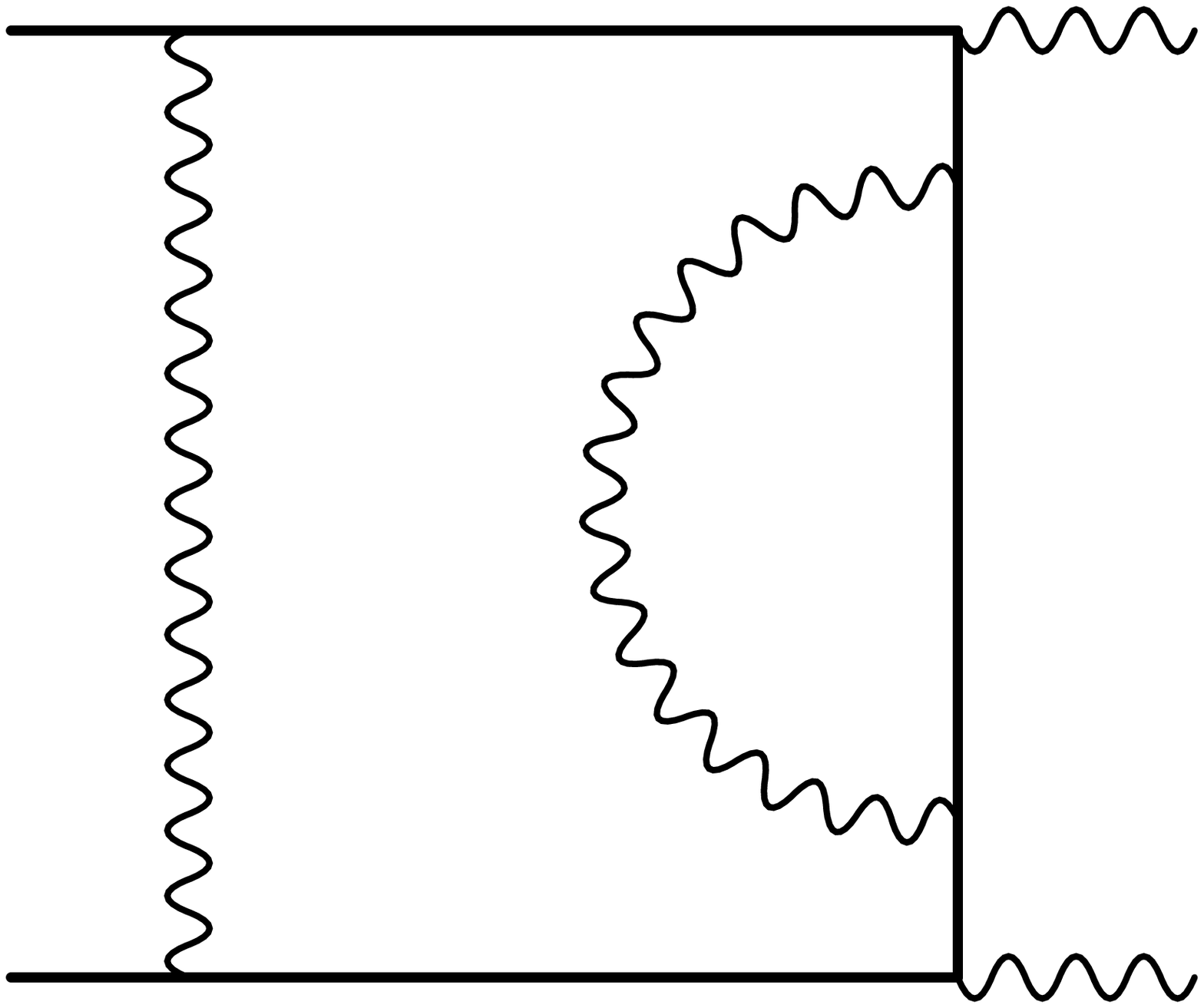,width=25mm}
&\hspace*{2mm}
\psfig{figure=  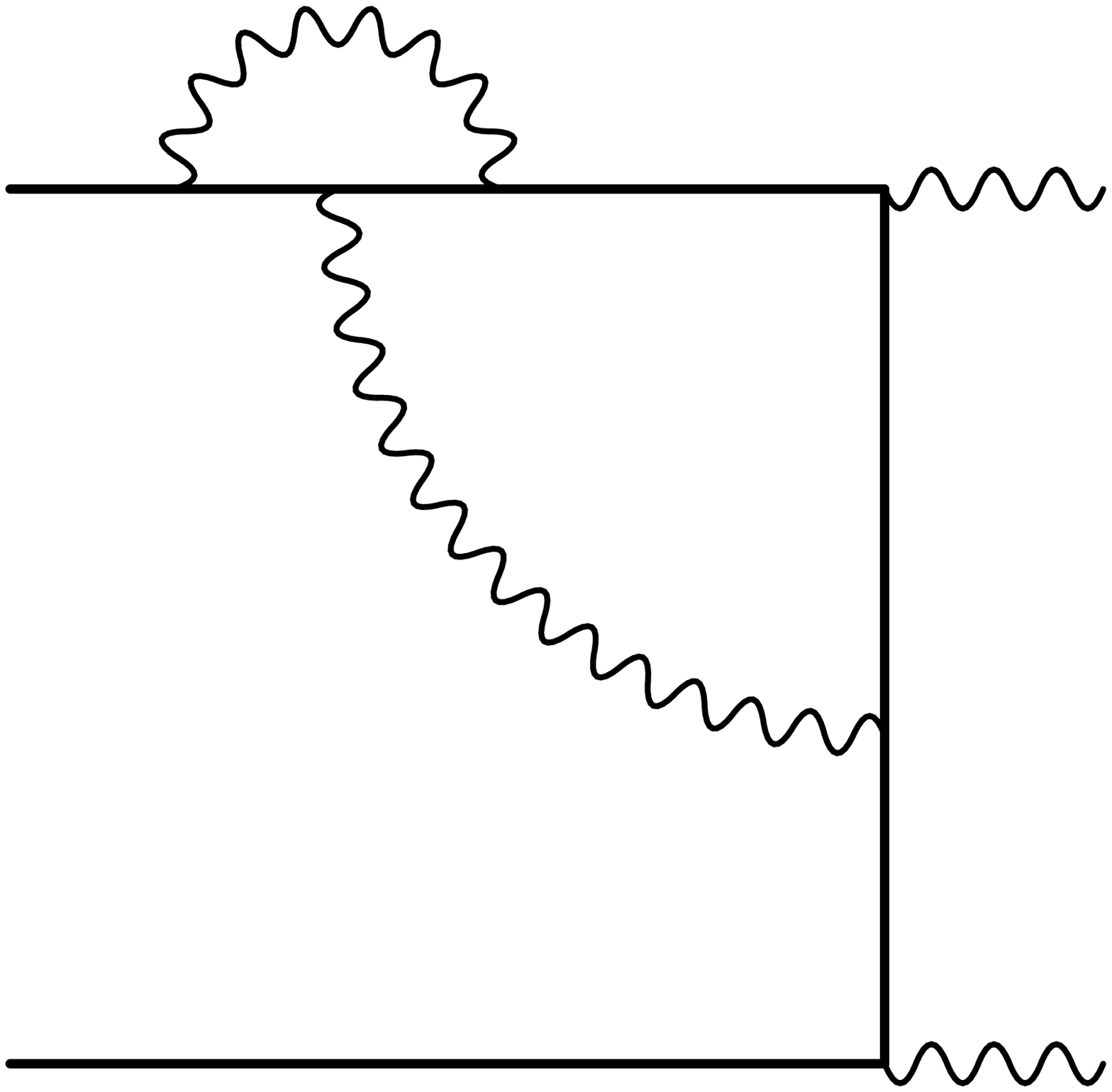,width=25mm}
&\hspace*{2mm}
\psfig{figure=  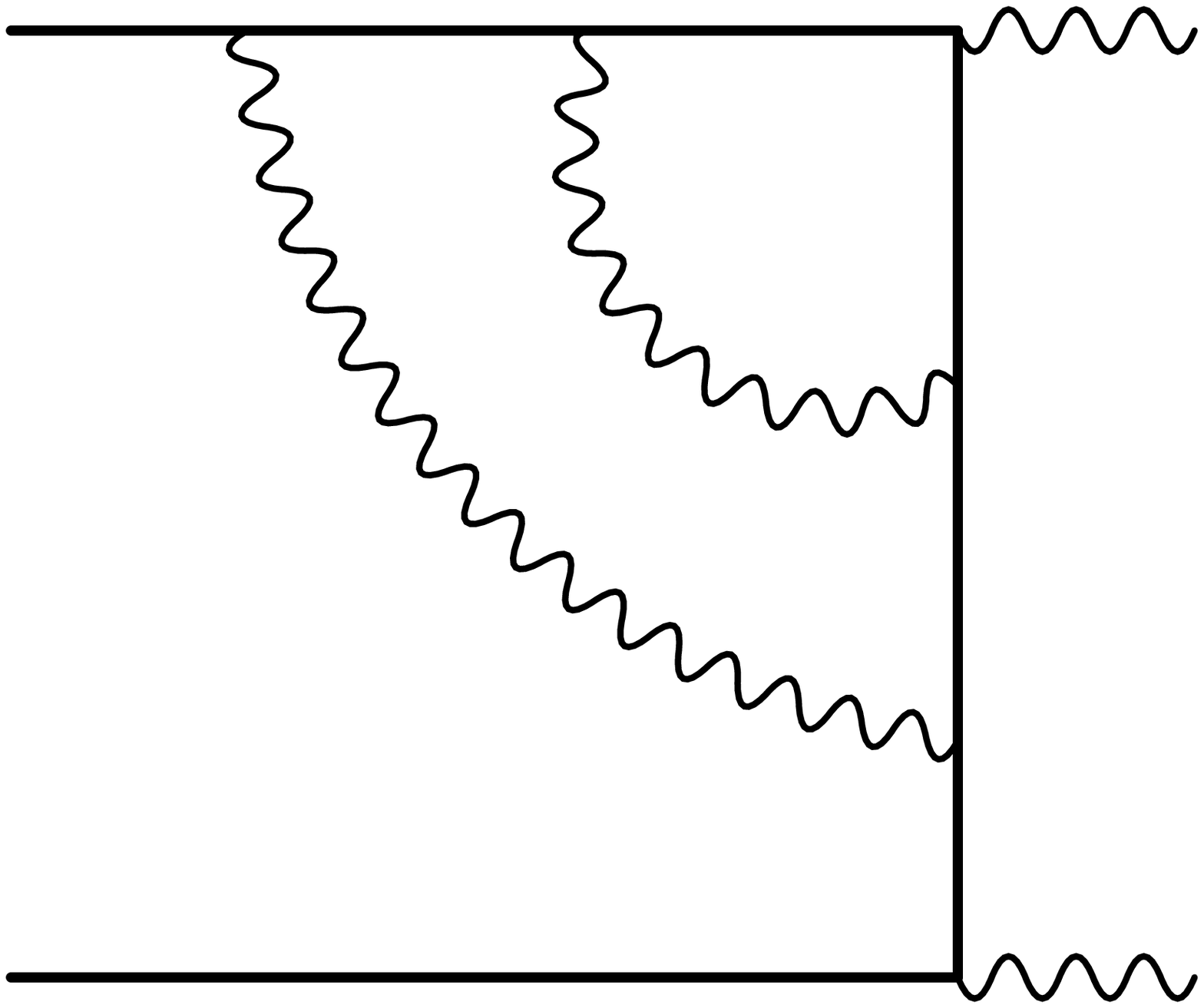,width=25mm}
&\hspace*{2mm}
\psfig{figure=  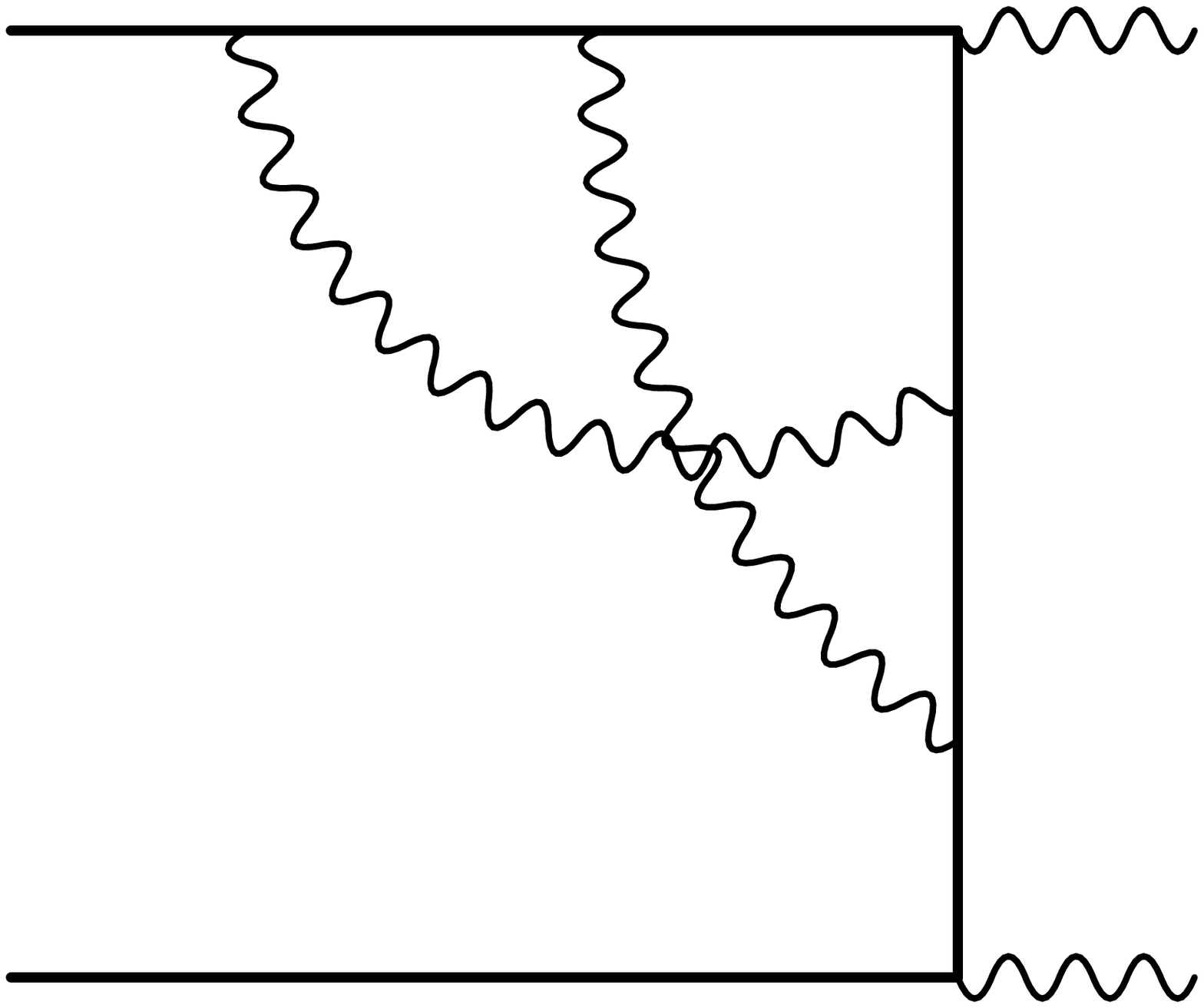,width=25mm}
\\[1mm]
  $D_6$ & $D_7$ & $D_8$ & $D_9$ & $D_{10}$
\\[1mm]
\psfig{figure=  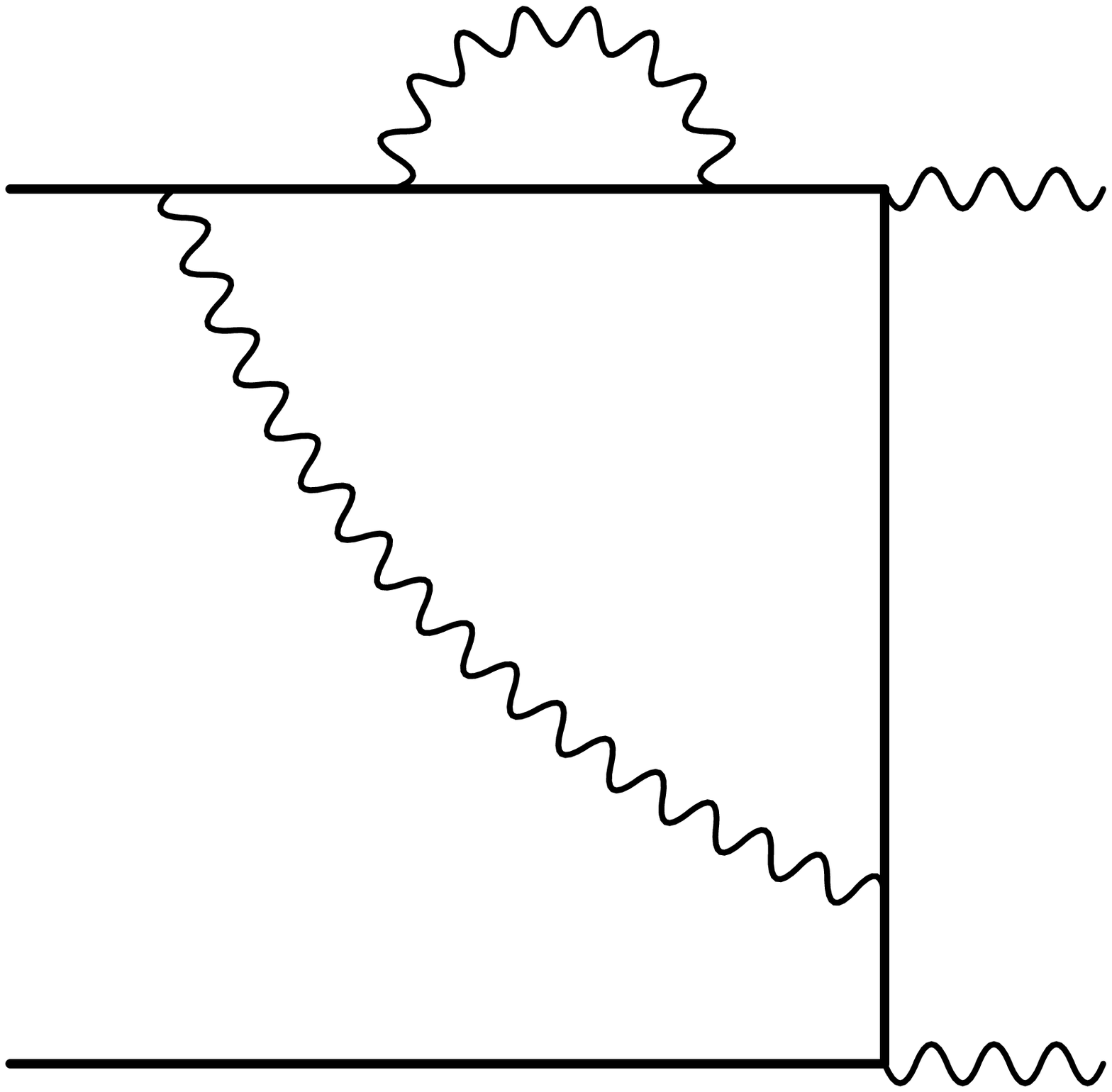,width=25mm}
&\hspace*{2mm}
\psfig{figure=  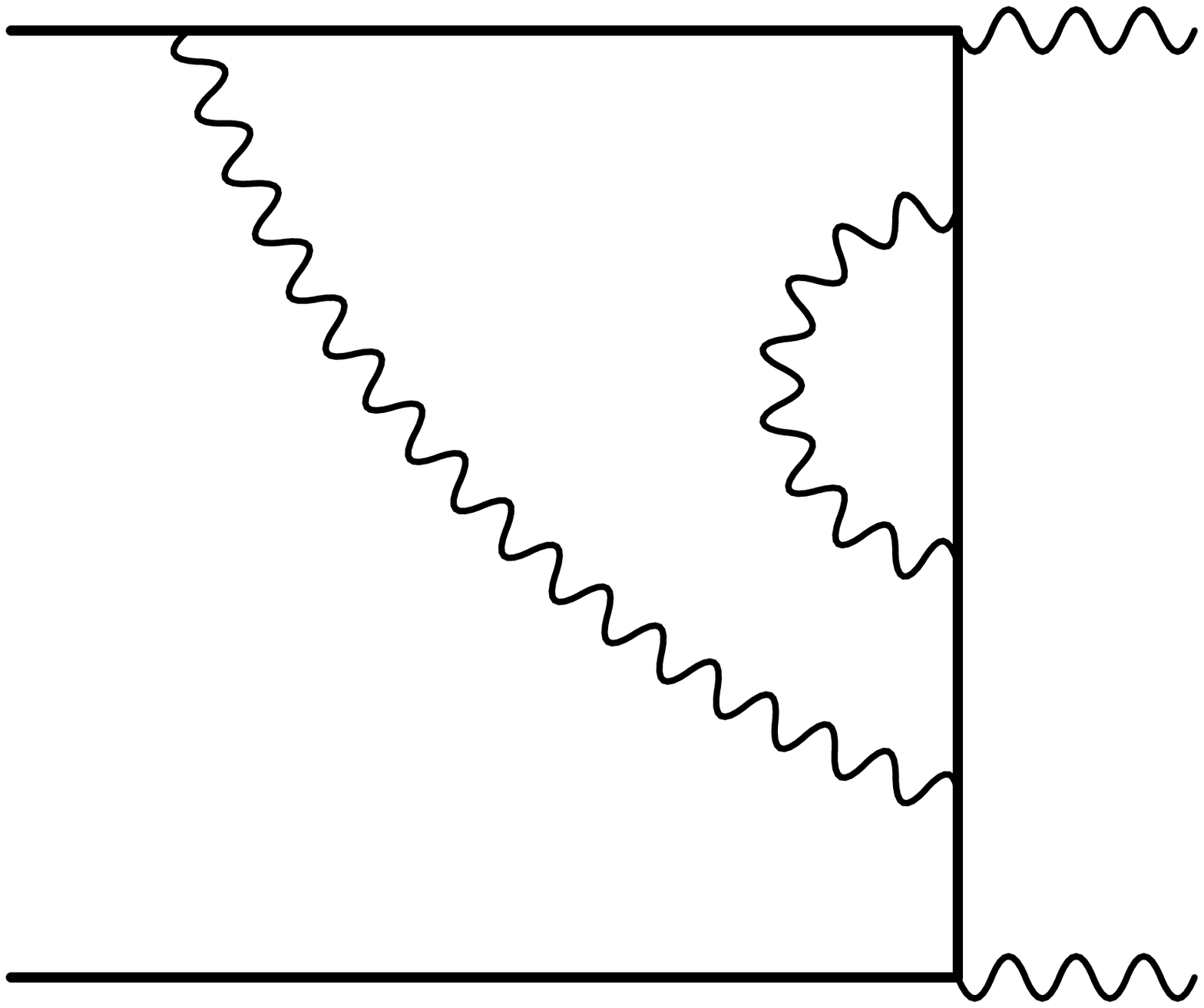,width=25mm}
&\hspace*{2mm}
\psfig{figure=  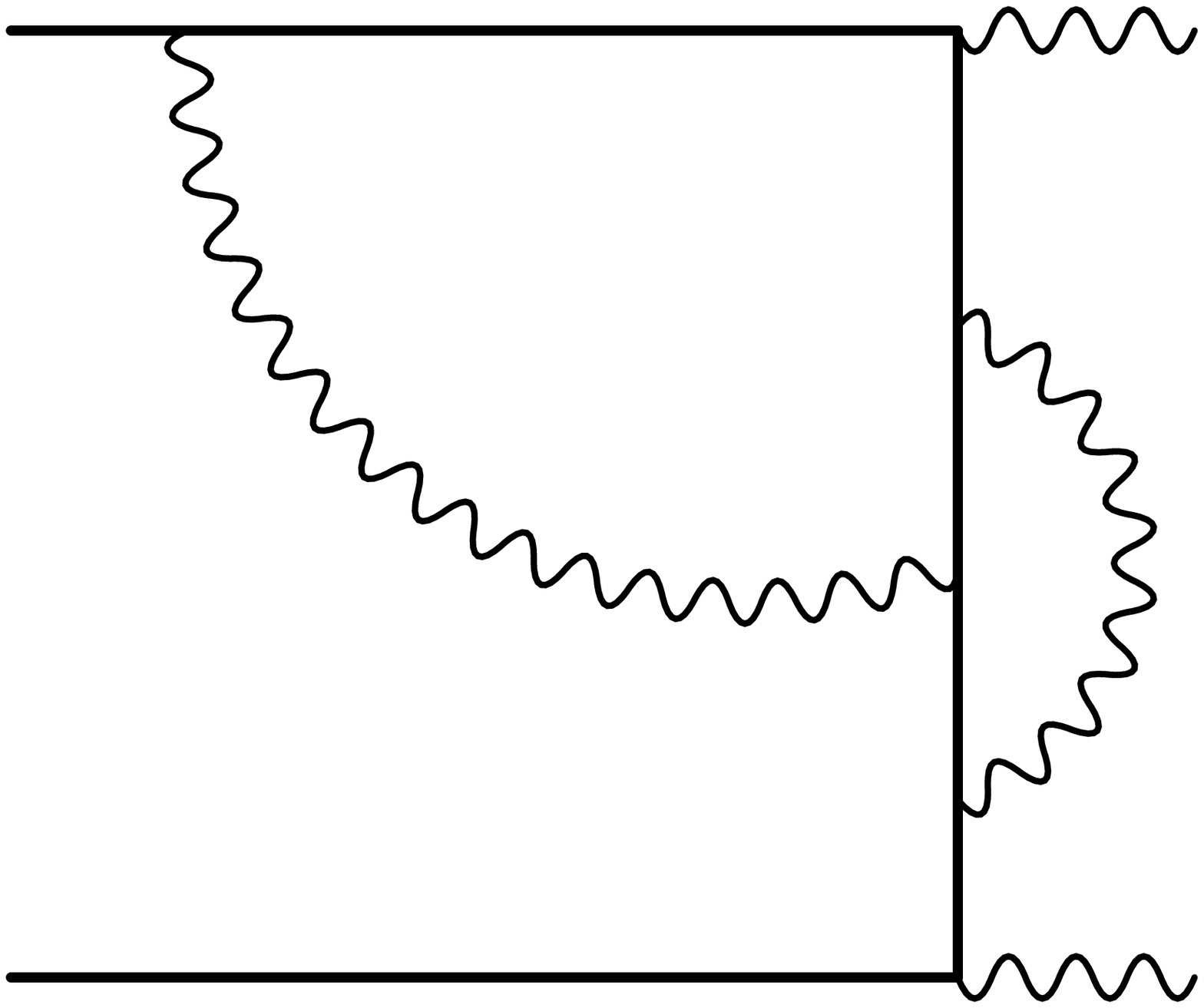,width=25mm}
&\hspace*{2mm}
\psfig{figure=  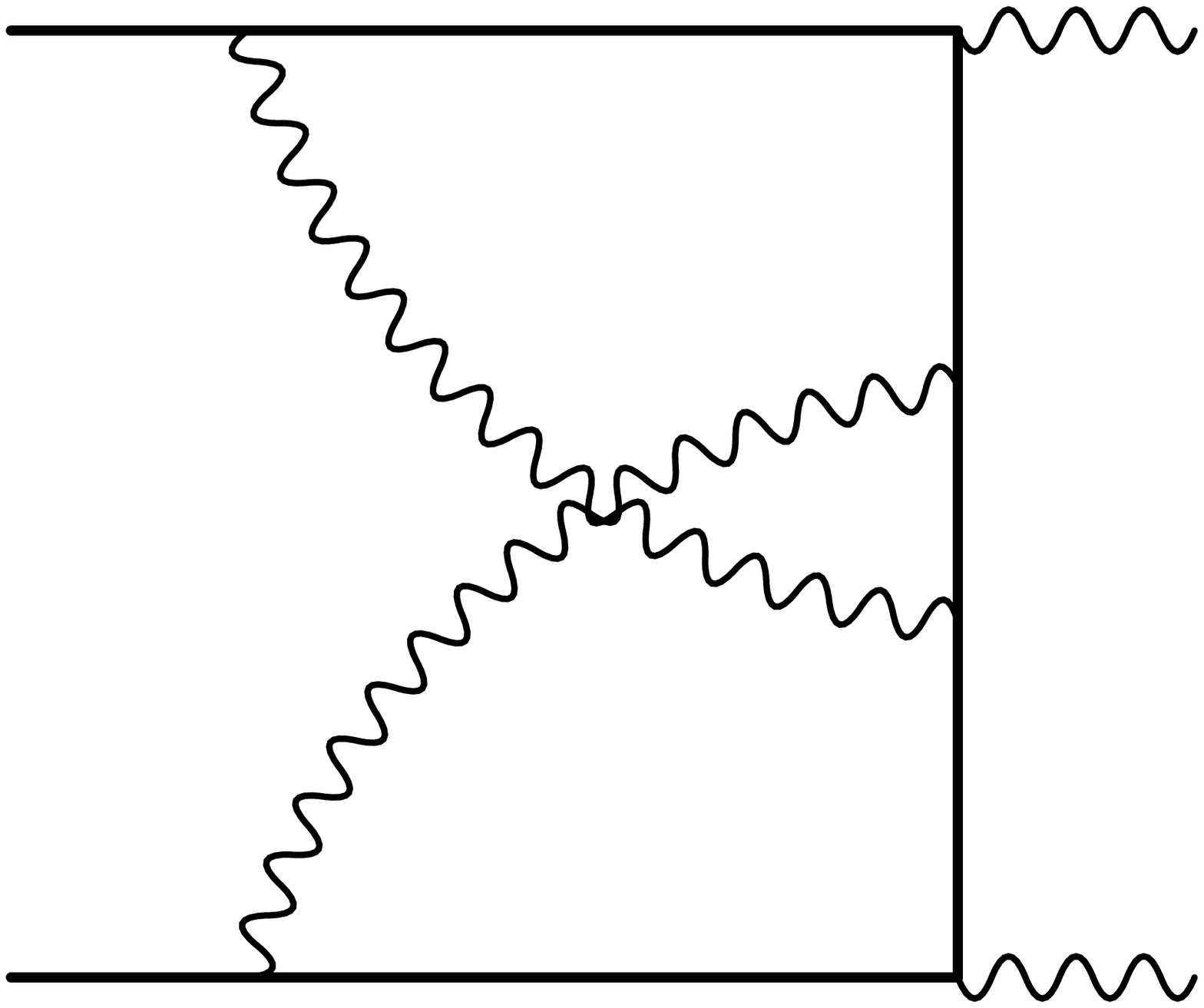,width=25mm}
&\hspace*{2mm}
\psfig{figure=  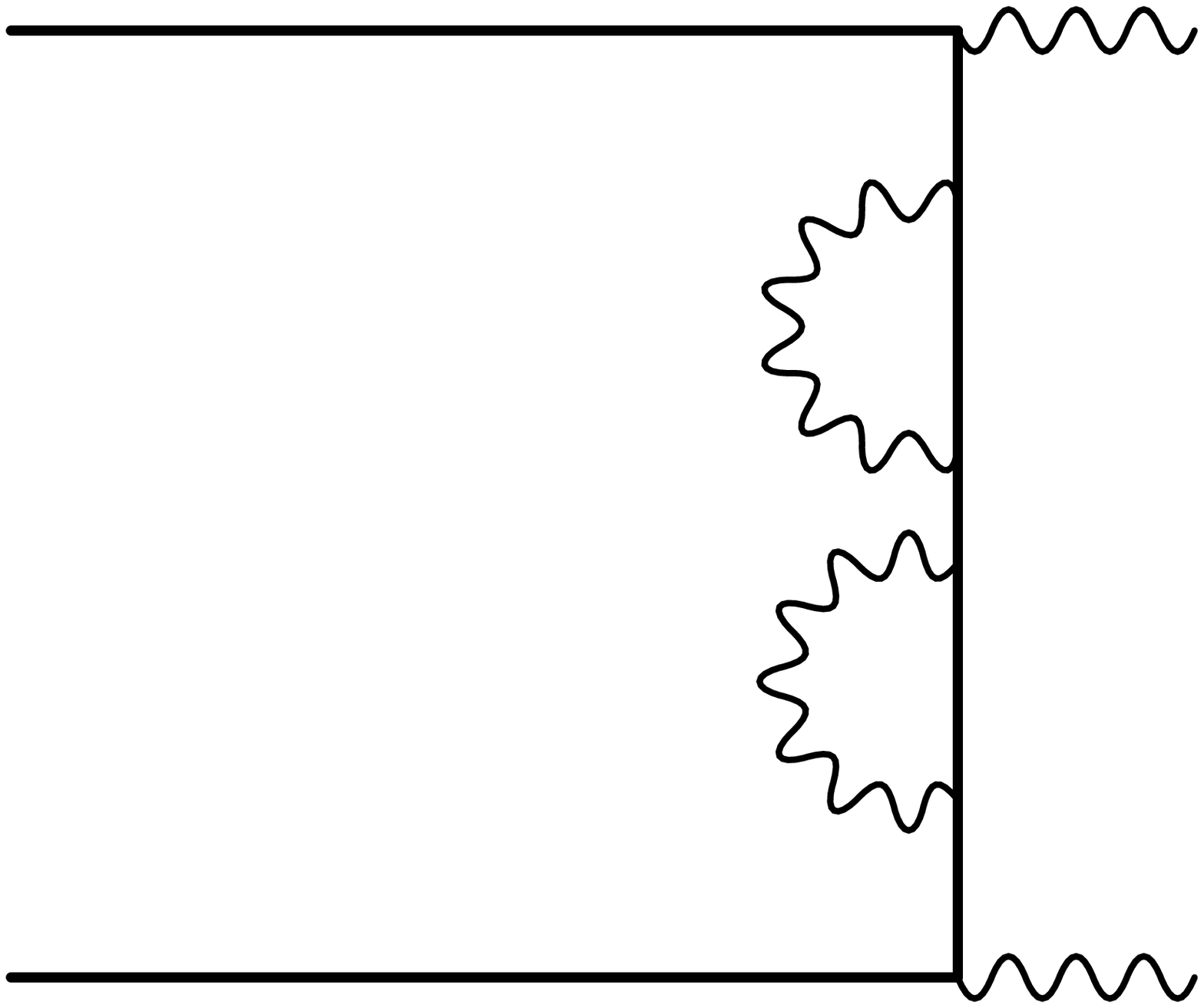,width=25mm}
\\[1mm]
 $D_{11}$ & $D_{12}$ & $D_{13}$ & $D_{14}$ & $D_{15}$
\\[1mm]
\psfig{figure=  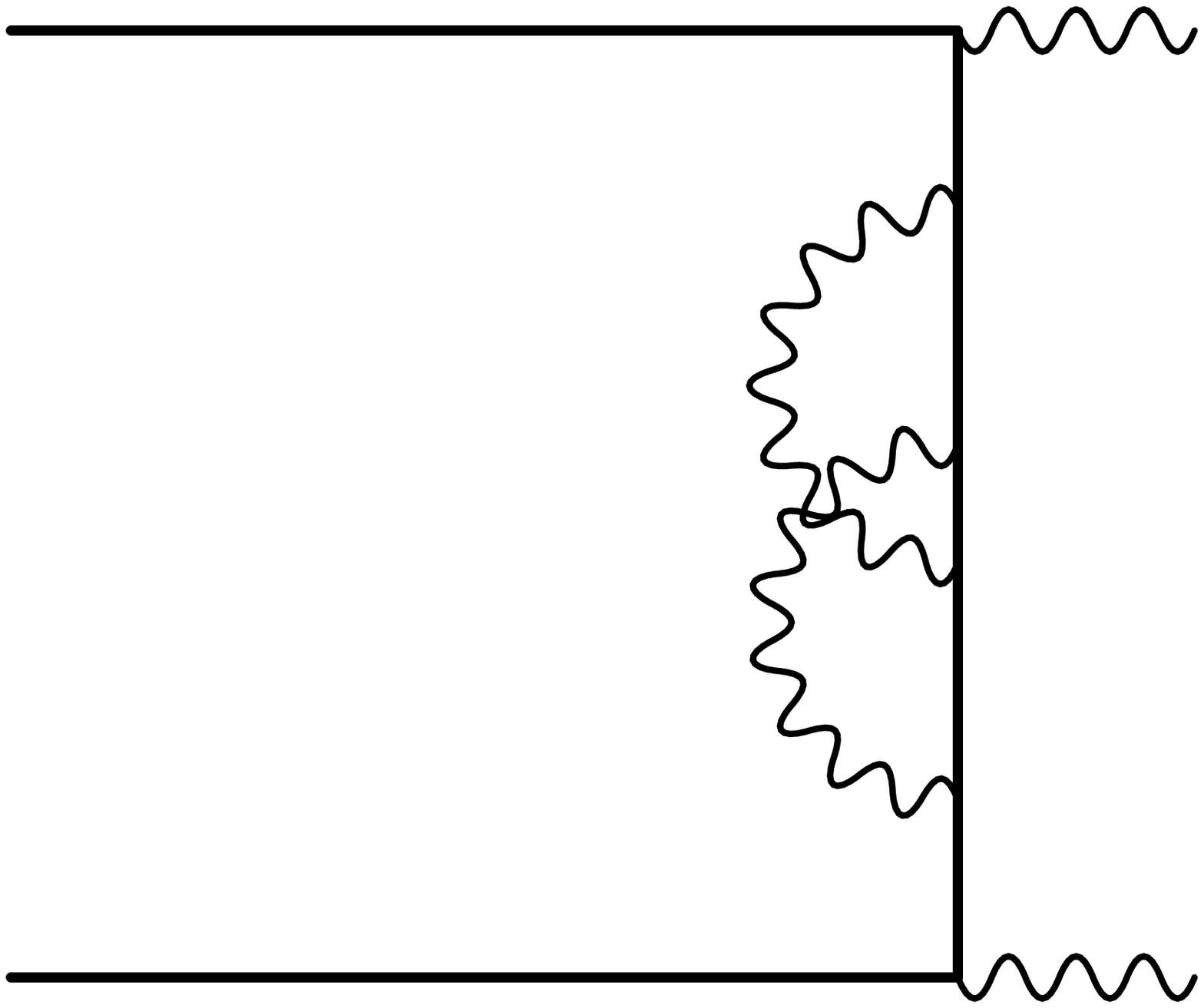,width=25mm}
&\hspace*{2mm}
\psfig{figure=  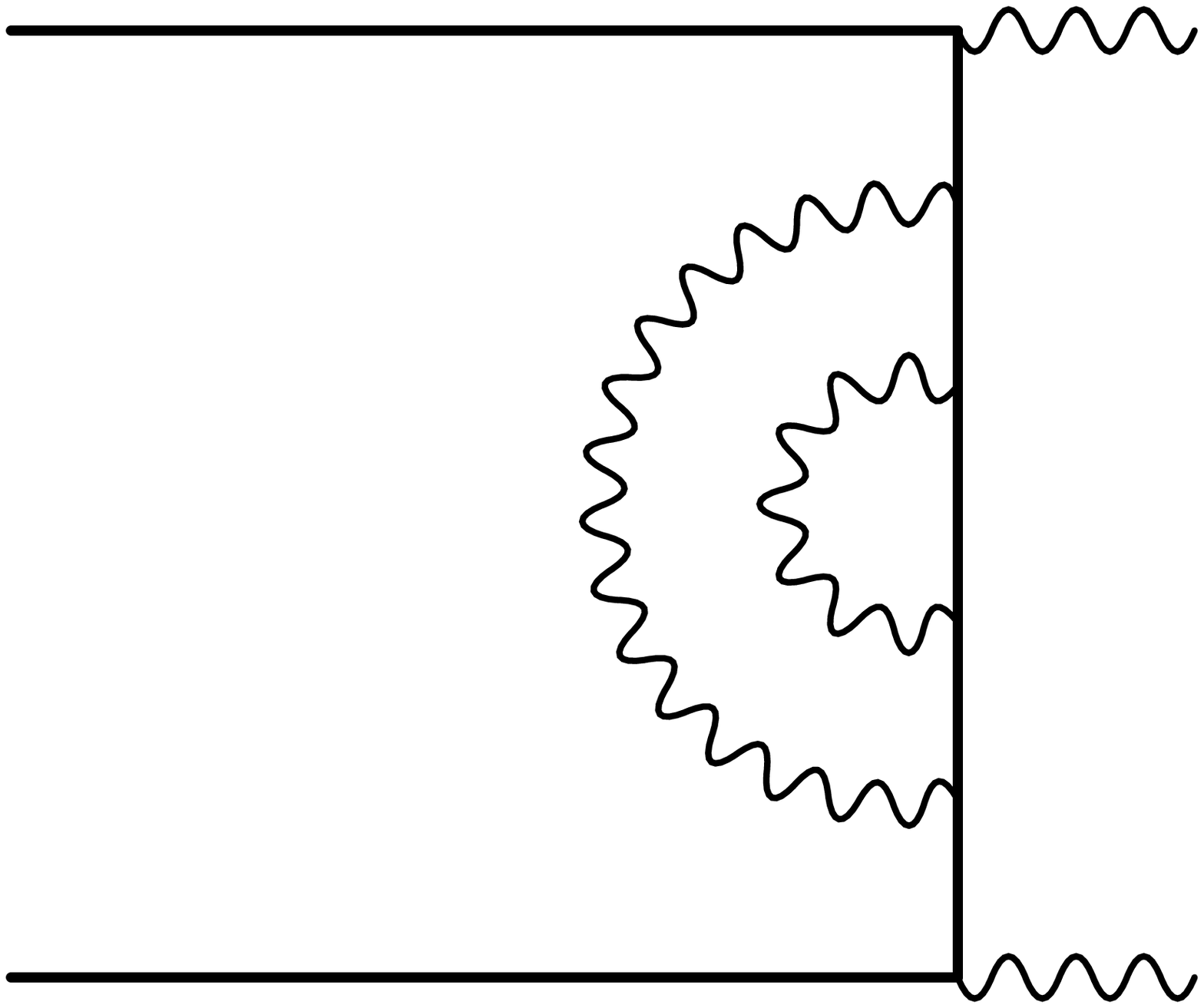,width=25mm}
&\hspace*{2mm}
\psfig{figure=  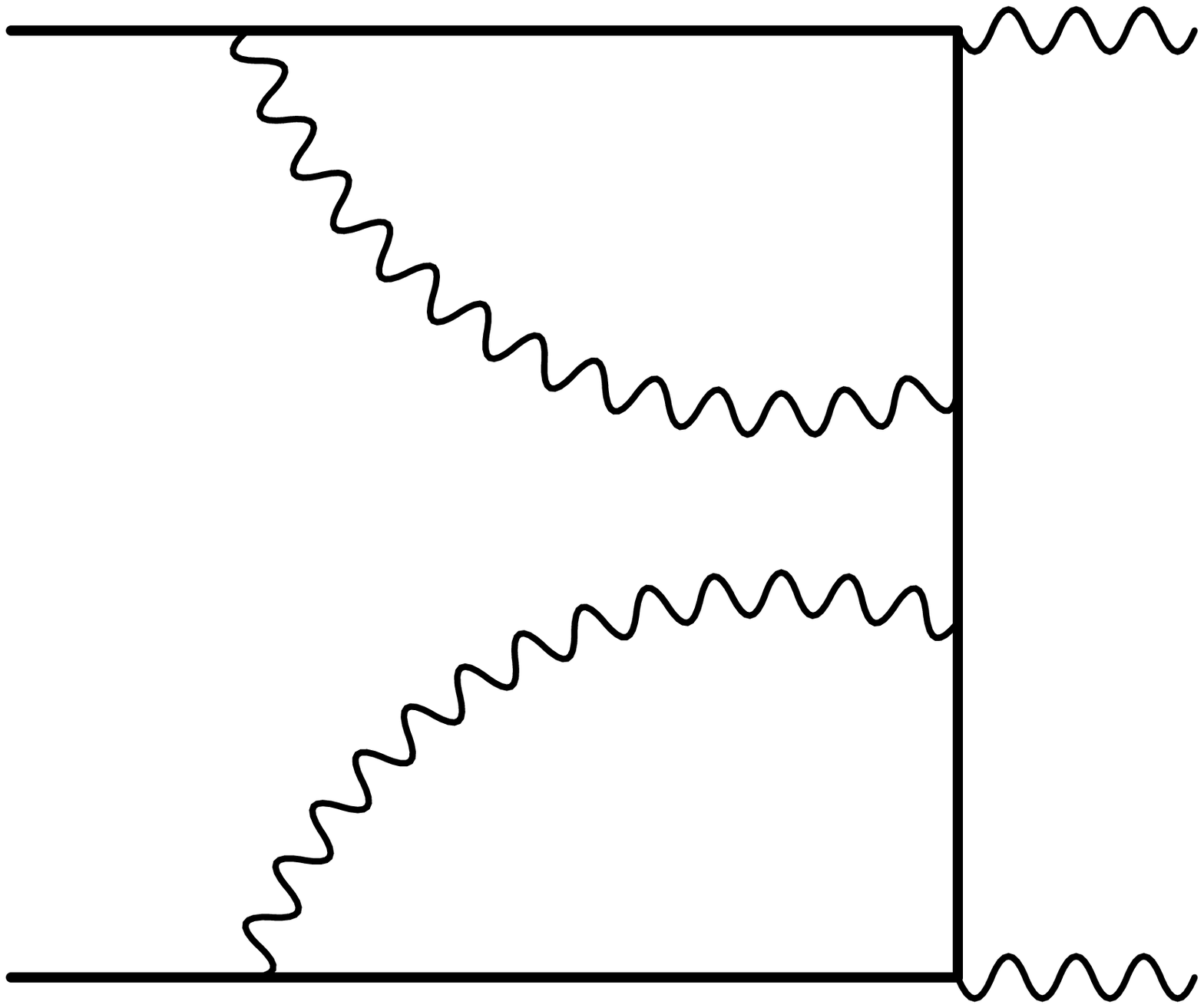,width=25mm}
&\hspace*{2mm}
\psfig{figure=  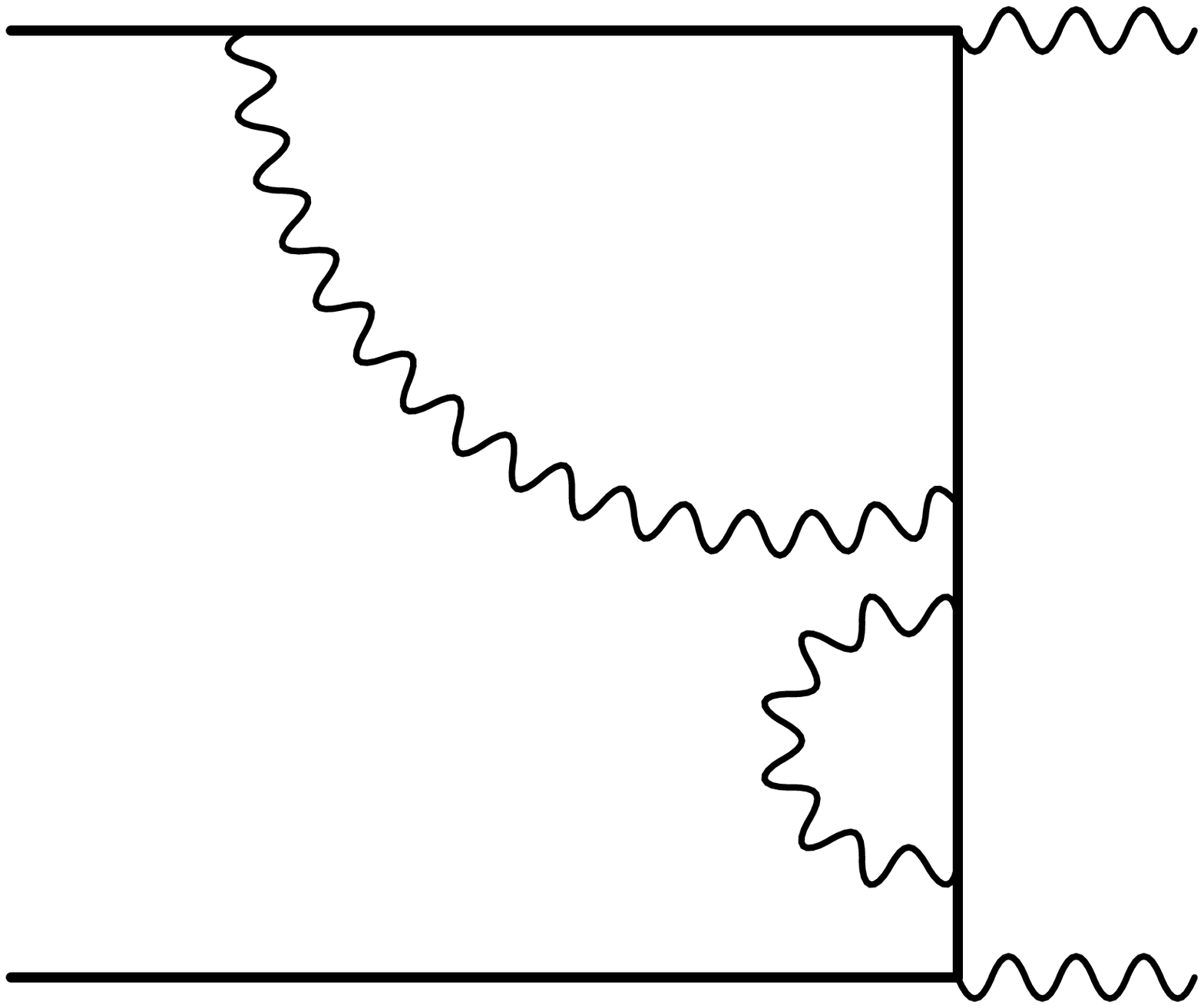,width=25mm}
&
\\[1mm]
$D_{16}$ & $D_{17}$ & $D_{18}$ & $D_{19 }$
&
\\[1mm]
\end{tabular}
}
\]
\end{minipage}
%\vspace*{10mm}
\caption{Two-loop photonic diagrams for the p-Ps decay.}
\label{fig:twoPs}
\end{figure}

\begin{figure}[h]
\hspace*{-2mm}
\begin{minipage}{16.cm}
\[
\mbox{
\begin{tabular}{cccc}
\psfig{figure=  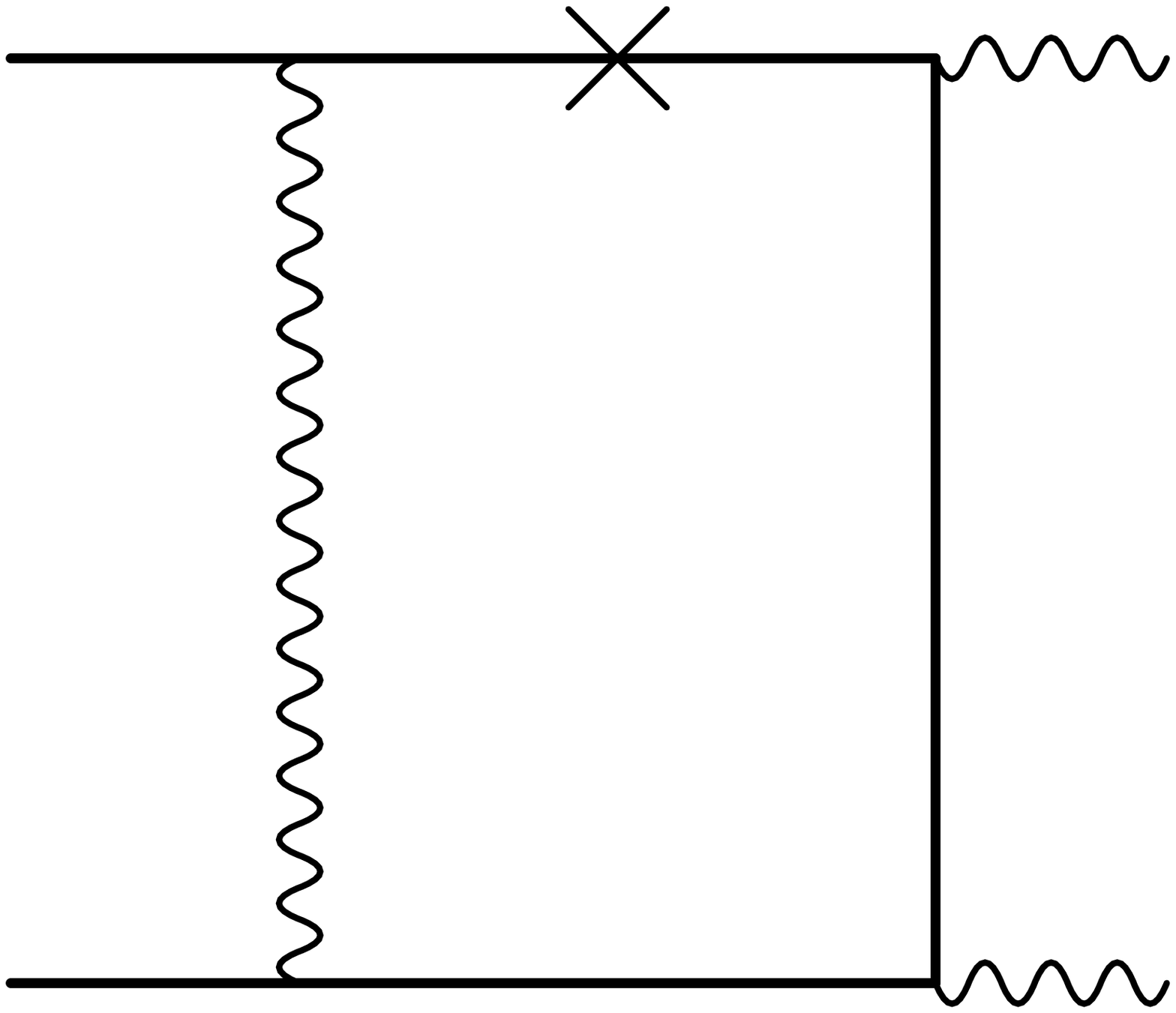,width=25mm}
&\hspace*{2mm}
\psfig{figure=  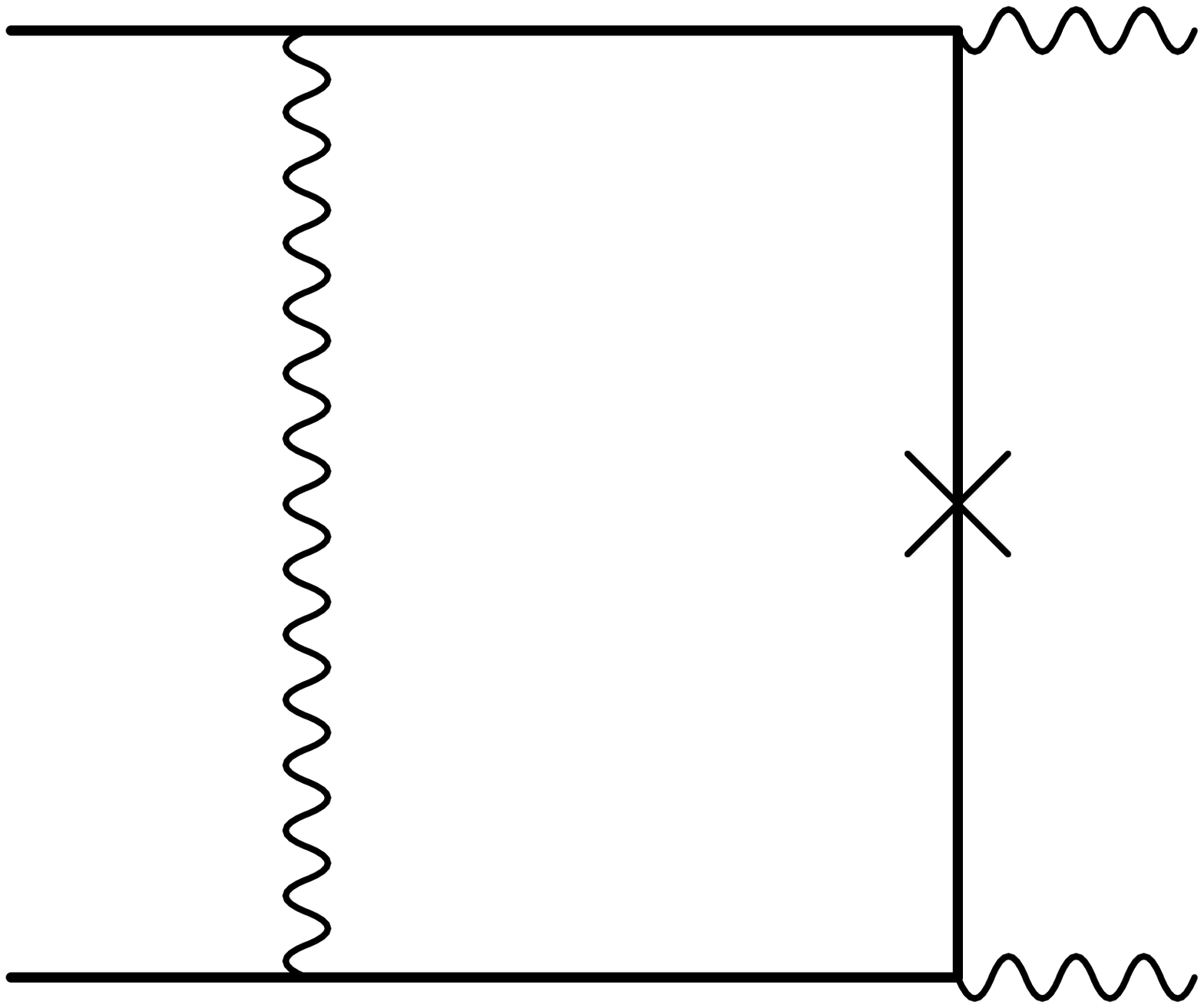,width=25mm}
&\hspace*{2mm}
\psfig{figure=  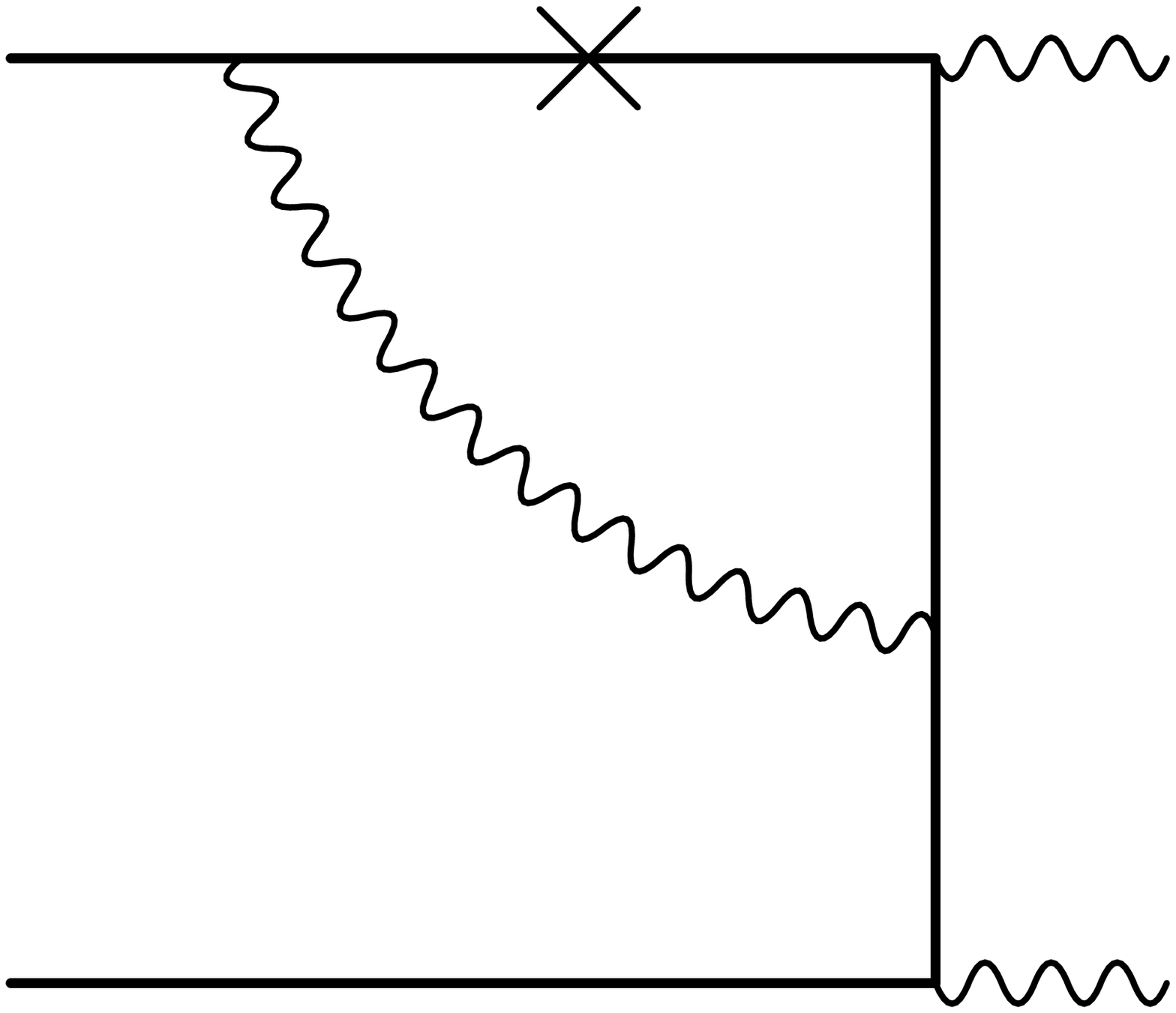,width=25mm}
&\hspace*{2mm}
\psfig{figure=  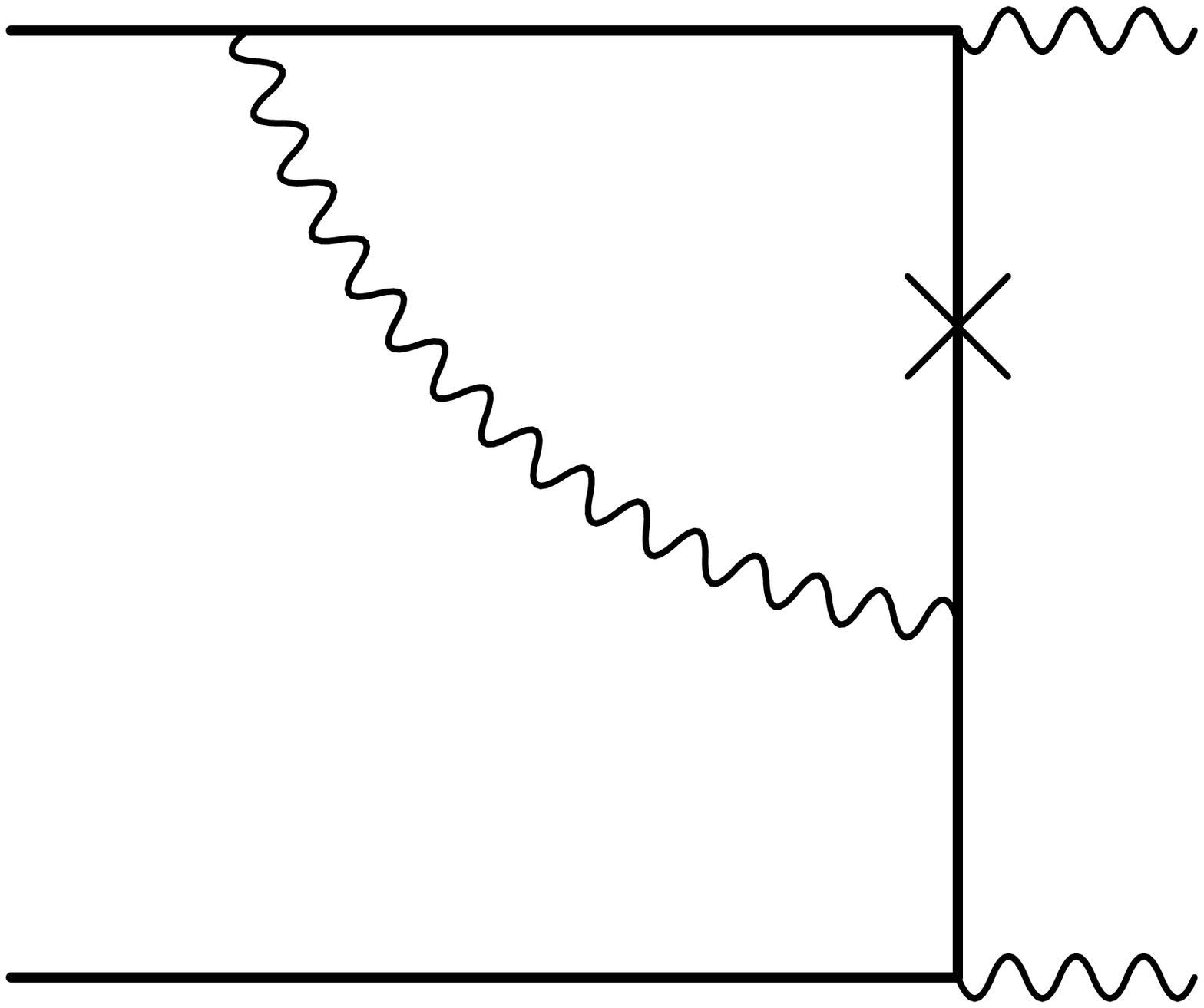,width=25mm}
\\[1mm]
 $C_5$ & $C_{7}$ & $C_{11}$  & $C_{12}$
\\[1mm]
%%%
\psfig{figure=  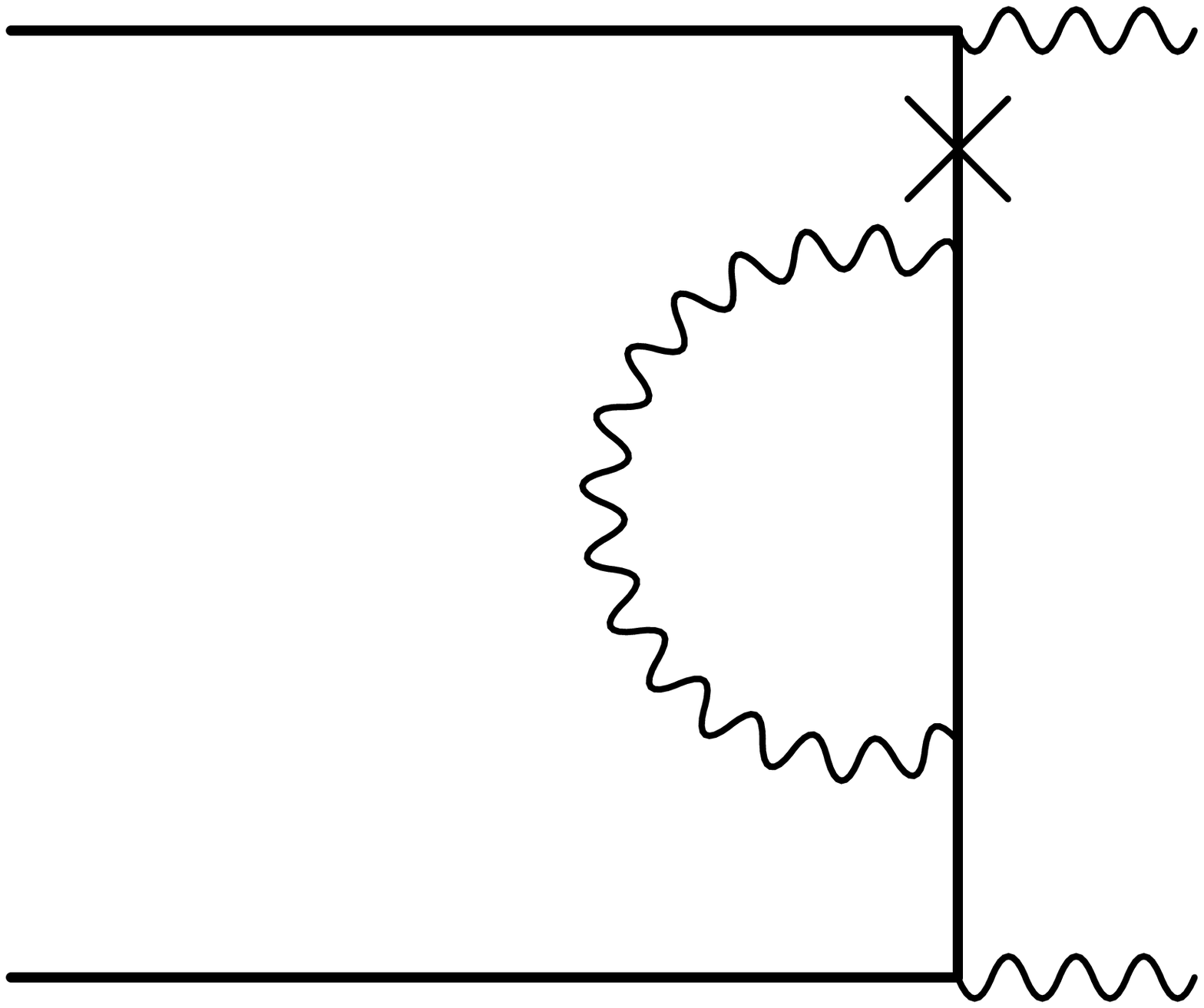,width=25mm}
&\hspace*{2mm}
\psfig{figure=  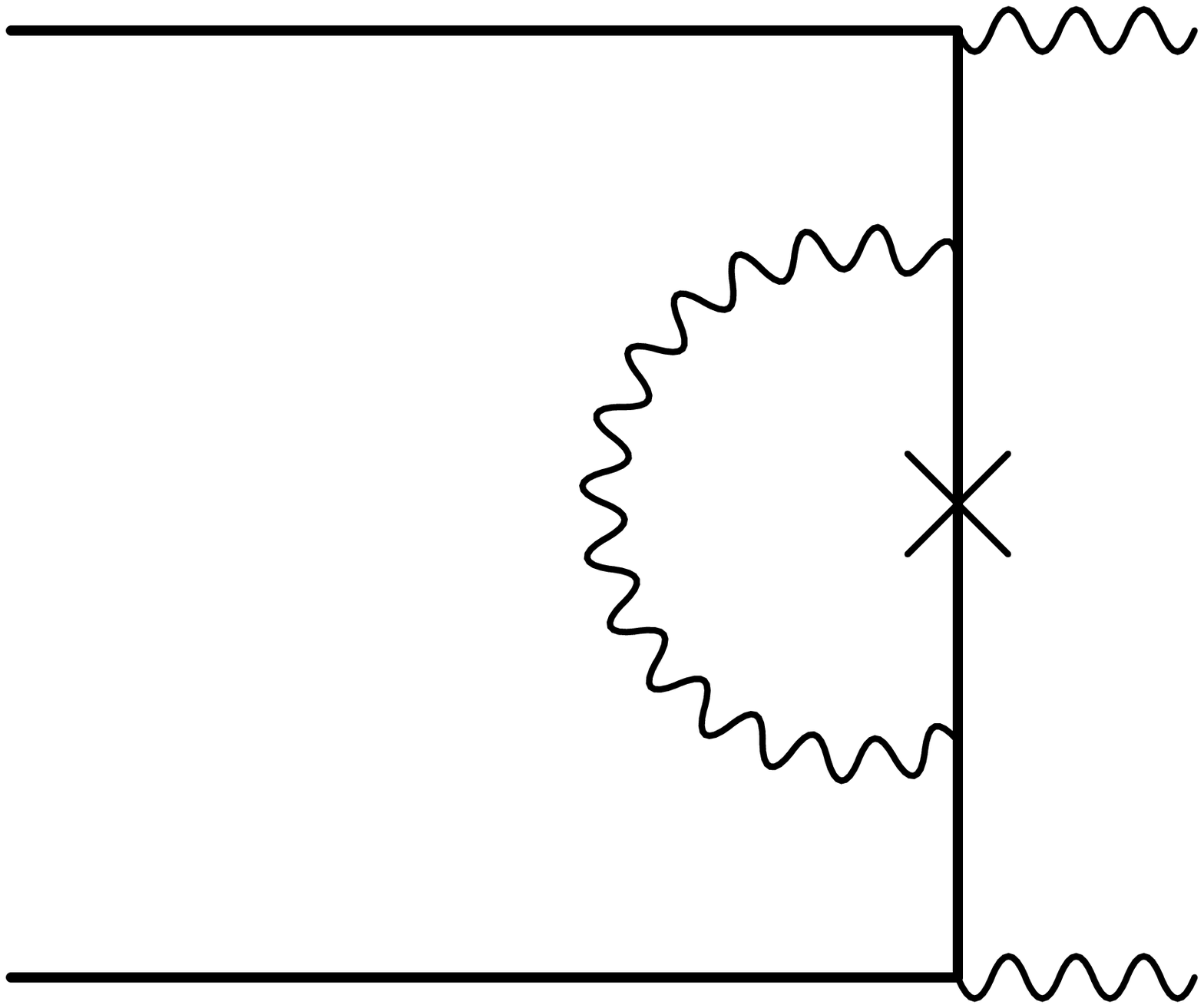,width=25mm}
&\hspace*{2mm}
\psfig{figure=  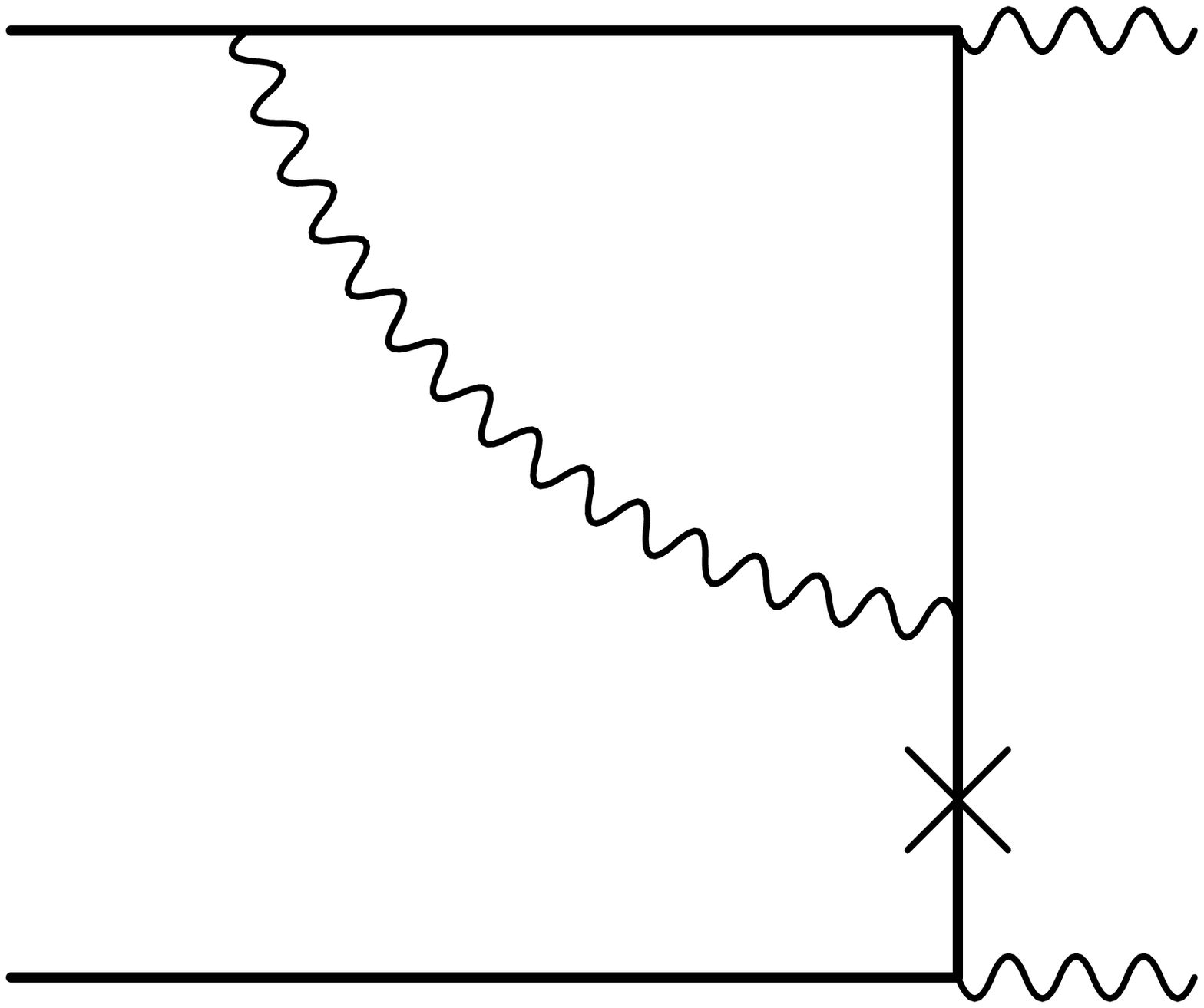,width=25mm}
&\hspace*{2mm}
\\[1mm]
$C_{15}$& $C_{17}$ & $C_{19}$ &
\\[1mm]
\end{tabular}
}
\]
\end{minipage}
%\vspace*{10mm}
\caption{Mass counterterms in one-loop diagrams. The numbering
conforms to Fig.~\protect\ref{fig:twoPs}.}
\label{fig:twoCT}
\end{figure}

Another class of corrections is generated by the two-photon diagrams,
shown in Fig.~\ref{fig:twoPs}.  These are the most difficult diagrams
we have to compute, since in general they diverge and a regularization
is required.  There arise two types of divergences: First, there are
the ultraviolet (UV) divergences; they are removed by an appropriate
renormalization.  Second, there are infrared (threshold)
singularities.  They will remain in the final result for the hard
scale contributions and vanish only in the sum with the soft
contributions.

The evaluation of the hard corrections is made possible by
 a combination of the analytical and numerical
methods.  The idea is to construct an infrared safe
expression from divergent Feynman amplitudes by subtracting
appropriate counterterms, which can be computed analytically.  The
construction of the counterterms is based on
the following observation: in a given Feynman diagram the infrared
singularities appear when the loop momenta are small. In this
situation the propagator of the virtual electron in the $t$-channel
can be contracted to a point.  As the result, the infrared behavior of
those Feynman diagrams is identical to that of the
two-loop three point functions considered earlier in
\cite{threshold,Czarnecki:1999mw}.

We proceed in the following way: after constructing an infrared
finite expression, we combine propagators using Feynman
parameters; perform momentum integrations analytically; extract
ultraviolet (UV) divergences and integrate
numerically over typically 5 (in some cases 6) Feynman parameters in
the finite expressions. For the numerical integration we use
the adaptive Monte Carlo routine VEGAS \cite{Vegas}.

Let us illustrate the basic steps of the calculation by considering
as an example the non-planar box diagram $D_1$ shown in
Fig.~\ref{fig:twoPs}.
A power counting shows that this diagram is UV finite but
IR divergent.  To demonstrate how the IR counterterm is
constructed we consider a symbolic expression for this diagram
(after taking the trace over Dirac matrices):
\be
D_1 \sim \int \frac {{\rm d}^D l_1}{(2\pi)^D} \frac {{\rm d}^D l_2}{(2\pi)^D}
\frac {f^{(0)} + f^{(1)} + f^{(2)} + ....}
{l_1^2 l_2^2 (l_1^2+2pl_1)(l_2^2-2pl_2)(l_3^2+2pl_3)(l_3^2-2l_3p)
(l_3^2+2p_1l_3+2m^2)}.
\ee
Here $p=(m,0)$ is the four momentum of the incoming electron or
positron, $l_3 = l_1+l_2$ is the sum of the loop momenta $l_{1,2}$,
and $p_1 = p - q$, where $q$ is the
four-momentum of the outgoing photon. The quantities $f^{(i)}$ in
the numerator denote the uniform functions of the loop momenta:
\be
f^{(i)}(\lambda l_1, \lambda l_2) = \lambda^{i}f^{(i)}(l_1,l_2).
\ee
Only terms with $f^{(0)}$ and $f^{(1)}$
diverge in IR.  We use the following identity:
\be
D_1 \equiv D_1^{(i \ge 2)} + \left (  D_1^{(i=0,1)}
- \left [ D_1^{(i=0,1)} \right ]_{\rm ct}
\right )
+\left [ D_1^{(i=0,1)} \right ]_{\rm ct}.
\label{d1}
\ee
where the counterterm $\left [ D_1^{(i=0,1)} \right ]_{\rm ct}$
is obtained by expanding the propagator of the electron in the
$t$-channel in Taylor series in small loop momenta,
\ba
\left [ D_1^{(i=0,1)} \right ]_{\rm ct}
\sim &&\frac {1}{2m^2} \int \frac {{\rm d}^D l_1}{(2\pi)^D}
\frac {{\rm d}^D l_2}{(2\pi)^D}
\frac {f^{(0)} + f^{(1)}}
{l_1^2 l_2^2 (l_1^2+2pl_1)(l_2^2-2pl_2)(l_3^2+2pl_3)(l_3^2-2l_3p)}
\nonumber \\
&& \times
\left ( 1 - \frac {l_3^2 +2p_1l_3}{2m^2}  \right ).
\ea
Examining Eq.~(\ref{d1}) one recognizes that the first two terms
in that equation are finite, both in the UV and the IR, and
hence can be evaluated numerically. The last term,
$\left [ D_1^{(i=0,1)} \right ]_{\rm ct}$, is divergent.
Since the $t$-channel propagator
has been contracted to a point,  this term corresponds
to a three-point, rather than four-point Feynman amplitude.
Such integrals were computed in a previous study \cite{threshold} (see
also \cite{Czarnecki:1999mw}).
Using those results one can obtain the counterterm  $\left
[ D_1^{(i=0,1)} \right ]_{\rm ct}$  analytically:
\be
\left [ D_1^{(i=0,1)} \right ]_{\rm ct} \sim
\frac {1}{\ep} \left ( 2\pi^2-4 \right ) +
48 + 12 \pi^2 \ln 2 + 6\pi^2 + 42\zeta_3.
\ee
Similar procedure was applied to evaluate the remaining
Feynman diagrams $D_i$. In some cases, like e.g. for the planar
box diagram $D_2$, the overall subtraction is not sufficient
and a more sophisticated approach is required.
The results of the calculation are summarized
in Table \ref{tab:Di}.

Finally, we have to consider the one-loop diagrams.  We need the
results including terms ${\cal O}(\ep)$, because they will be
multiplied by divergent renormalization constants. The
results are summarized in Table \ref{tab:Si}.
%oneLoop}.

For the renormalization one needs the electron wave function
renormalization constant, $\delta Z_e \equiv Z_e-1$, and the mass
counterterm $\delta m$ computed in dimensional regularization to
${\cal O}(\alpha^2)$.  These results can be found in \cite{bgs91}.
For completeness, we collect here the relevant formulas:
\ba
\delta Z_e &=& a\; \delta Z_e^{(1)} + a^2 \;\delta Z_e^{(2)}
\nonumber \\
\frac {\delta m}{m} &=&
a \; {\delta m^{(1)}\over m} + a^2 \; {\delta m^{(2)}\over m},
\ea
where
\ba
a &=& \frac {e^2}{(4\pi)^{D/2}}m^{(-2\ep)},
\nonumber \\
\delta Z_e^{(1)} = {\delta m^{(1)}\over m}
  &=&  - {3\over \ep}-4 -\ep \left( {\pi^2\over 4} + 8 \right) ,
\nonumber \\
\delta Z_e^{(2)} &=&
 \frac {9}{2\ep^2}+\frac {51}{4\ep}-\frac {49 \pi^2}{4}
 +16 \pi^2 \ln 2-24 \zeta_3 +\frac {433}{8},
\nonumber \\
{\delta m^{(2)}\over m} &=&
\frac {9}{2\ep^2}+\frac {45}{4\ep}-\frac {17 \pi^2}{4}
 +8 \pi^2 \ln 2-12 \zeta_3 +\frac {199}{8}.
\ea

\begin{figure}[h]
\hspace*{-7mm}
\begin{minipage}{16.cm}
\[
\mbox{
\begin{tabular}{ccc}
\psfig{figure=  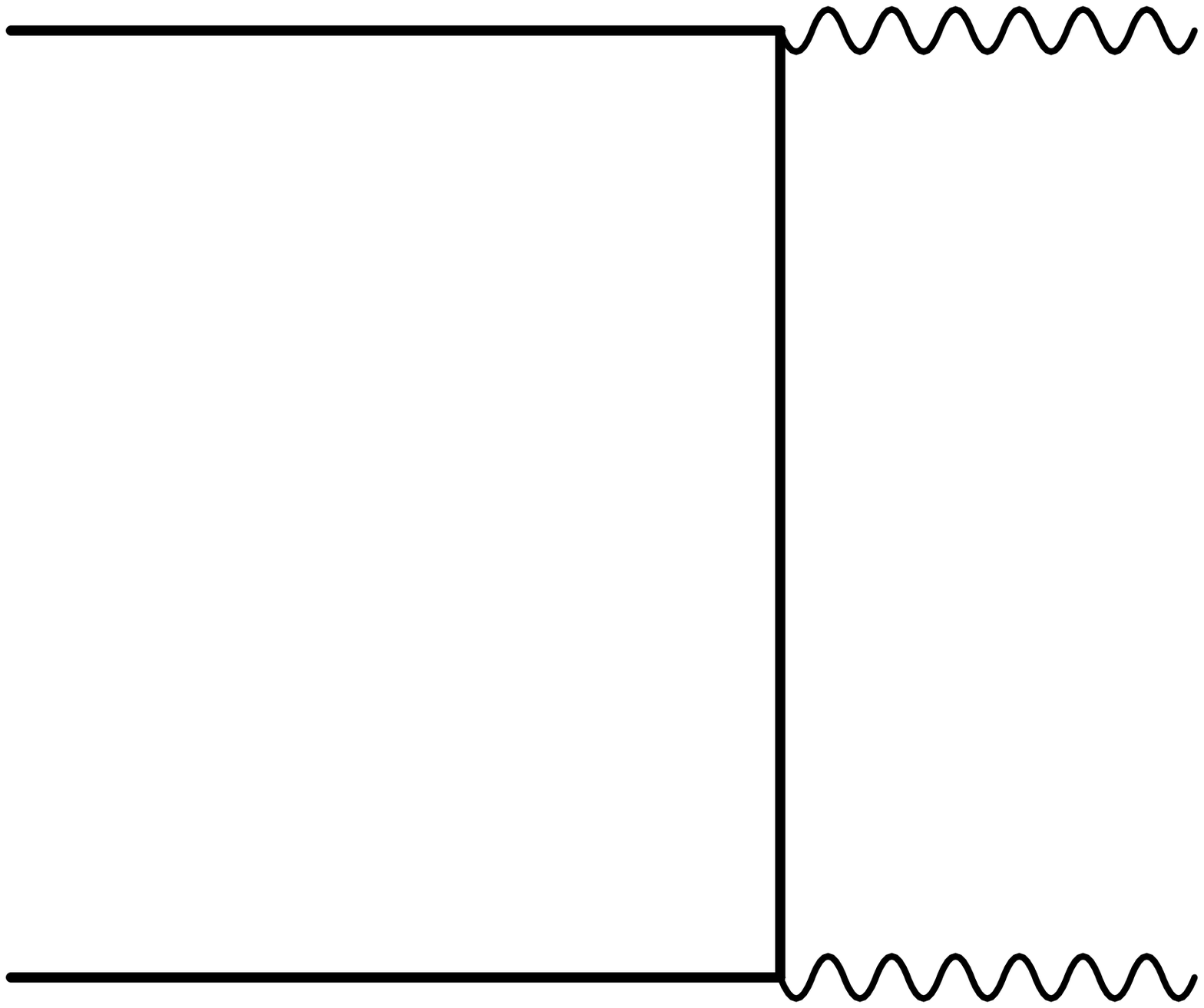,width=25mm}
&\hspace*{5mm}
\psfig{figure=  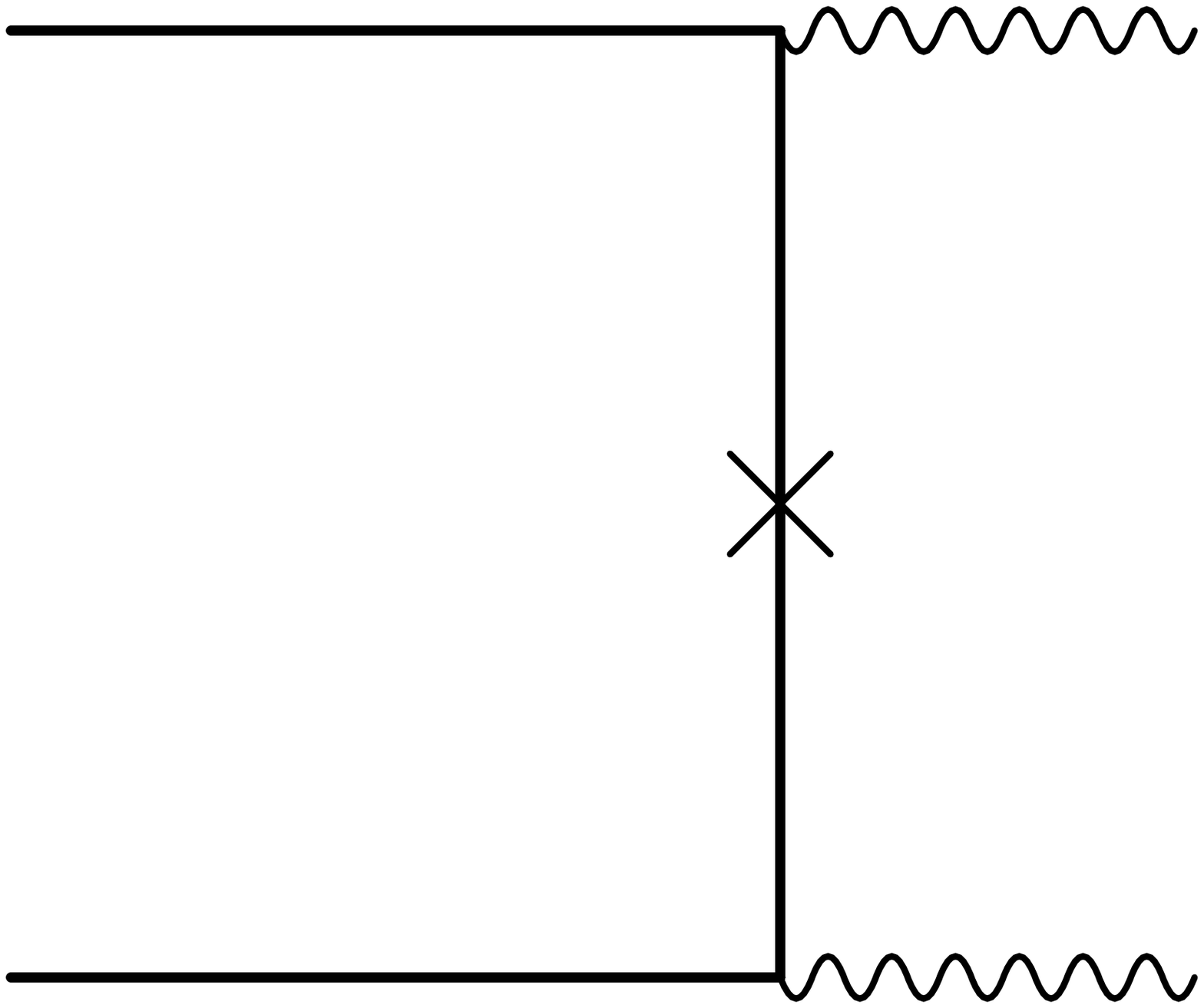,width=25mm}
&\hspace*{5mm}
\psfig{figure=  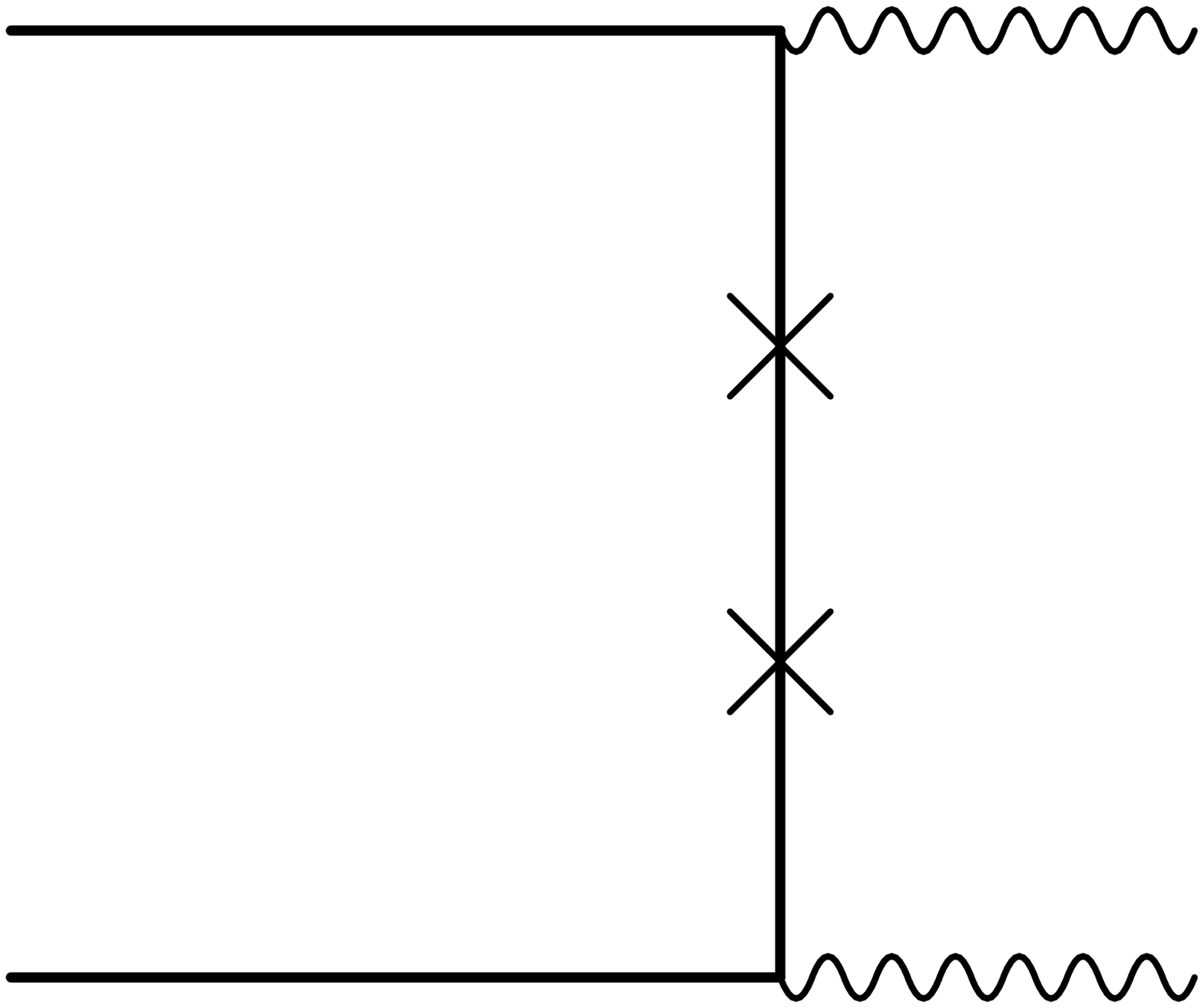,width=25mm}
\\[1mm]
 $B_0=1/8$ &\hspace*{5mm} $B_1=-1/8$ &\hspace*{5mm} $B_2=1/16$
\\[1mm]
\end{tabular}
}
\]
\end{minipage}
%\vspace*{10mm}
\caption{Tree-level amplitude of the p-Ps decay, higher order mass
counterterms, and their values.}
\label{fig:born}
\end{figure}

The final result is obtained by putting the many pieces together:
\ba
m^{4\epsilon} B_p^{\rm hard}(\gamma\gamma) &=&
 \delta Z_e^{(2)} B_0 + {\delta m^{(2)}\over m} \; B_1
+ \left( {\delta m^{(1)}\over m}\right)^2 B_2
\nonumber \\
&&
+ \delta Z_e^{(1)} \left( S_1+S_2+S_3+
                         {\delta m^{(1)}\over m} \; B_1 \right)
\nonumber \\
&&
+ {\delta m^{(1)}\over m} \; \left( \sum_{i=5,7,11,12,15,17,19} C_i\right)
\;+\; \sum_{i=1}^{19} D_i
\nonumber \\
&=& -{\pi^2\over 2\vep}-42.19(27).
\label{eq:4219}
\ea
Here $B_i$ are the tree-level diagrams (Fig.~\ref{fig:born}), $S_i$
are one-loop diagrams (Table \ref{tab:Si}), $C_i$ are the one-loop
diagrams with mass insertions (Table \ref{tab:Ci}), and $D_i$ are the
two-loop diagrams (Table \ref{tab:Di}).

For the complete hard correction we add Eqs.~(\ref{VP}, \ref{lbl},
\ref{eq:4219}) and find
\be
B_p^{\rm hard} = - {\pi^2\over 2\epsilon} +2\pi^2\ln m -40.46(30).
\label{hard}
\ee

\section{Final result}
\label{secFinal}
The final result for the second order non-logarithmic correction to
the p-Ps decay rate into two photons is obtained as a sum of the soft
(\ref{soft}) and hard (\ref{hard}) pieces, and the square of one-loop
corrections (\ref{oneloop2}).  Adding them one finds
\be
B_p = 5.1(3),
\label{final}
\ee
and the theoretical prediction for the p-Ps lifetimes becomes 
\begin{eqnarray}
\Gamma_{\rm p-Ps}^{\rm theory} &=& 7989.64(2)\,\mu {\rm s}^{-1}.
\label{finalresult}
\end{eqnarray}
In this equation we have not included the contribution of the decay
mode p-Ps$\to 4 \gamma$.  This decay channel increases $\Gamma_{\rm
p-Ps}$ by approximately $0.01\,\mu {\rm s}^{-1}$ \cite{AdkBr,Lepage:1983yy}.

\section{Conclusion}
\label{secConc}
We have described the numerical and analytical methods employed in our
calculation of the second order QED corrections to the p-Ps lifetime.
Our final result, the non-logarithmic correction ${\cal
O}(\alpha^2/\pi^2)$ with the coefficient $B_p = 5.1(3)$ increases
the decay rate by approximately $0.1~\mu {\rm s}^{-1}$.  The resulting
theoretical prediction is in excellent agreement with experiment.

The framework of this calculation is the Nonrelativistic QED with
dimensional regularization. The dimensional regularization facilitates
the separation of scales, the cornerstone of the effective theory. Its
main technical advantage is that no additional scales, such as a
photon mass, are introduced by the regularization, so that only single
scale integrals need to be computed. Unfortunately, these integrals
are still too complicated to be evaluated analytically.  We computed
them numerically by subtracting IR counterterms.  Those were
constructed using simpler integrals which are known analytically, as
we explained in Section \ref{sec4}.  The
multidimensional finite integrals were computed using adaptive Monte
Carlo integration routine VEGAS \cite{Vegas}.

The approach described in this paper can be also applied to the
calculation of the ${\cal O}(\alpha^2)$ corrections to the o-Ps decay
into three photons. In particular, the IR counterterms can be
constructed in a similar manner.  However, the larger number of diagrams
and two additional integrations over the three-photon phase space make
this problem significantly more difficult.

Somewhat surprising is the smallness of the second order corrections
found in this paper.  It resulted from a  strong cancellation
between soft and hard pieces computed in dimensional
regularization. We would like to stress that the soft and hard pieces
are not separately finite and depend on the regularization (in this
sense they are ``scheme-dependent'').  For this reason, large
constants accompanying divergent pieces in
Eqs.~(\ref{hard},\ref{soft}) may have no direct physical meaning.
Unambiguous information is provided by the scheme-independent results,
Eqs.~(\ref{oneloop2},\ref{ga},\ref{VP},\ref{lbl}), which are all of
order $[\rm several~units] \times (\alpha/\pi)^2$.

In the absence of a complete result on ${\cal O}(\alpha^2)$ to
$\Gamma_{\rm o-Ps}$ it is interesting to discuss what our result for
p-Ps might imply for the o-Ps lifetime puzzle. Although nothing can be
said rigorously, we believe that our result indicates that no dramatic
enhancement in the ${\cal O}(\alpha^2)$ effects in o-Ps decay is
possible.  In Table \ref{tab:comp} we have summarized available
results on the second order effects for both p-Ps and o-Ps decays (we
have not included the partial results for the soft corrections in o-Ps
since they are scheme-dependent).
One can see that, with the exception of $B^{\rm squared}$, all
radiative corrections are comparable for o-Ps and p-Ps decays. On the
other hand, significantly larger value of $B^{\rm squared}$ for o-Ps
can be traced back to a larger value of the one-loop correction to
o-Ps$\to 3 \gamma$ rate. The relation of the one-loop corrections for
p-Ps and o-Ps decay, however, is rather natural since the number of
diagrams is  approximately three times  larger  for o-Ps
decay. Thus, apart from the factor related to the number of
Feynman diagrams, there seems to be no significant difference in the
structure of radiative corrections to o-Ps and p-Ps decays.

Therefore, it is difficult to imagine that a complete
calculation of the ${\cal O}(\alpha^2)$ correction
to o-Ps decay will result in a dramatically large number,
necessary to resolve the o-Ps lifetime puzzle.
We believe that this puzzle will be solved by continuing experimental
studies and  we look forward to learning their results.

\section{Acknowledgments}
We are grateful to G.~Adkins, R.~Fell, and J.~Sapirstein for pointing
out an error in our initial evaluation of the light-by-light
contributions (\ref{l12}). 
This research was supported in part by the United States Department of
Energy under grants DE-AC02-98CH10886 and DE-AC03-76SF00515, by BMBF
under grant BMBF-057KA92P, by Gra\-duier\-ten\-kolleg
``Teil\-chen\-phy\-sik'' at the University of Karlsruhe, by the
Russian Foundation for Basic Research under grant 99-02-17135, and by
the Russian Ministry of Higher Education.

%\bibliographystyle{../../pro/tex/revtex}
%\bibliography{../../pro/tex/phd}

%%%%%%%%%%%%%%%%%%%%%%%%%%%%%%%%%%%%%%%%%%%%%%%%%%%%%%%%%%%%%%%%%%%%%%

%%%%%%%%%%%%%%%%%%%%%%%%%%%%%%%%%%%%%%%%%%%%%%%%%%%%%%%%%%%%%%%%%%%%%%
\begin{table}
\caption{Values of one-loop diagrams}
\label{tab:Si}
\begin{center}
\begin{tabular}{crrr}
 Diagram & $\ep^{-1}$ & $\ep^0$ & $\ep^1$ \\
\hline
$S_1$ & ${1\over 4}$ & $ -{1\over 2}$ &
      $1 + {5\over 2}\ln 2 + {\pi^2\over 48}$
\\
$S_2$ & ${1\over 4}$ & ${\pi^2\over 16}-\ln 2$  &
      $-2\ln 2 +\ln^2 2 + {\pi^2\over 16} + {7\over 8}\zeta_3$
\\
$S_3$ & $-{1\over 2}$ & $-{3\over 4}+\ln 2$  &
      $-{3\over 2} + {3\over 2}\ln 2 - \ln^2 2 + {\pi^2\over 24}$
\\
\hline
Total & 0 &  $-{5\over 4} + {\pi^2\over 16}$  &
 $-{1\over 2} + 2\ln 2 +{\pi^2\over 8} + {7\over 8}\zeta_3$  \\
\end{tabular}
\end{center}
\end{table}

\begin{table}
\caption{Values of counterterm diagrams}
\label{tab:Ci}
\begin{center}
\begin{tabular}{crrr}
 Diagram & $\ep^{-1}$ & $\ep^0$ & $\ep^1$ \\
\hline
$C_5$ &
       $ {1 \over 4} $
&
    $   - {1 \over 3} + {1 \over 2} \ln 2$
&
       $ {37 \over 18} + {3 \over 8} \pi^2 + {1 \over 2} \ln 2 - {1
       \over 2} \ln^2 2 $ \\
$C_7$ &
       $  - {1 \over 4} $
&
     $  {1 \over 2} + {1 \over 2} \ln 2$
&
 $  - 1 - {7 \over 24} \pi^2 + {5 \over 2} \ln 2 - {1 \over 2} \ln^2 2$
\\
$C_{11}$ &0
&
       $ 1 - {1 \over 16} \pi^2 - \ln 2$
&
       $ 3 - {7 \over 48} \pi^2 - {5 \over 2} \ln 2 + \ln^2 2 - {7
       \over 8} \zeta_3 $
\\
$C_{12}$ &0
&
      $ - 1 + {3 \over 16} \pi^2 - \ln 2$
&
       $  - 3 - {1 \over 48} \pi^2 + {1 \over 2} \ln 2 + \ln^2 2 + {21
       \over 8} \zeta_3 $
\\
$C_{15}$ &
       $ {5 \over 8} $
&
       $  {3 \over 4} - \ln 2$
&
       $ {3 \over 2} - {1 \over 32} \pi^2 - {3 \over 2} \ln 2 + \ln^2
       2 $
\\
$C_{17}$ &
       $  - {1 \over 2} $
&
       $ - {3 \over 4} + 2 \ln 2$
&
       $  - {3 \over 2} + {1 \over 8} \pi^2 + \ln 2 - 2 \ln^2 2 $ \\
$C_{19}$ &
       $  - {1 \over 4} $
&
     $  - {1 \over 16} \pi^2 + {1 \over 2} \ln 2$
&
       $  - {1 \over 24} \pi^2 + {1 \over 2} \ln 2 - {1 \over 2} \ln^2
       2 - {7 \over 8} \zeta_3 $
\\
\hline
Total & $-{1\over 8}$ & ${1\over 6}+{\pi^2\over 16}+{1\over 2}\ln 2$
 & ${19\over 18}-{\pi^2\over 32}+ \ln 2 - {1\over 2}\ln ^2 2 +{7\over
    8} \zeta_3 \simeq 2.2519$
\end{tabular}
\end{center}
\end{table}

\begin{table}
\caption{Values of two-photon diagrams in Fig.~\protect\ref{fig:twoPs}.}
\label{tab:Di}
\begin{center}
\begin{tabular}{crrr}
 Diagram & $\ep^{-2}$ & $\ep^{-1}$ & $\ep^0$ \\
\hline
$D_1$ & 0 & $-\frac{1}{2} + \frac{\pi^2}{4}$ & 37.35(10) \\
$D_2$ & 1 & $1$                              & $-43.69(20)$ \\
$D_3$ & 0 &  0                               & $-0.074(2)$  \\
$D_4$ & ${1\over 4}$ &  $-\frac{1}{2}-\frac{3\,\pi^2}{4}$
                     & $-65.90(3)$\\
$D_5$ & $-{1\over 4}$ &  $-1 + \frac{3}{2}\ln 2$ & 57.918(20) \\
$D_6$ & ${1\over 2}$ & $-1 - 2\,\ln 2 + \frac{\pi^2}{8}$ &  9.661(5)
\\
$D_7$ & $-1$ & ${1\over 2}+{7\over 2}\ln 2$  & $-10.20(1)$ \\
$D_8$ & ${1\over 8}$ & ${3\over 16}-\ln 2+{\pi^2\over 16}$
                     & $-0.324$ \\
$D_9$ &  ${1\over 8}$ & ${5\over 16}-\ln 2+{\pi^2\over 16}$
                     & $1.475(50)$ \\
$D_{10}$ & 0 & $-{1\over 2}$ & $3.488(2)$ \\
$D_{11}$ & $-{1\over 8}$ & ${45\over 16}-{\pi^2\over 4}-2\ln 2$
                     & $-3.80(3)$ \\
$D_{12}$ & $-{1\over 8}$ & $-{51\over 16}+{\pi^2\over 2}-2\ln 2$
         & $1.69(12)$ \\
$D_{13}$ & ${1\over 8}$ & ${3\over 16}+{\pi^2\over 16}-\ln 2$
         & $-0.12(1)$ \\
$D_{14}$ & 0 & 0 & $-0.804(2)$ \\
$D_{15}$ & ${23\over 16}$ & $-5\ln 2 +{15\over 4}$ &
  $\frac{39}{4} - \frac{17\,\pi^2}{96} -
  \frac{27\,\ln 2}{2} + 9\,\ln^2 2 \simeq 2.9689$ \\
$D_{16}$ & $-{1\over 2}$ & $2\ln 2-{3\over 4}$ & $-1.885(4)$ \\
$D_{17}$ & $-{1\over 2}$ & $5\ln 2-{9\over 4}$ & $-1.175(1)$ \\
$D_{18}$ & ${1\over 8}$ & ${\pi^2\over 16}-\ln 2$ &
 $ \frac{\pi^2}{16} + \frac{\pi^4}{128} -
  2\,\ln 2 - \frac{\pi^2\,\ln 2}{4} +
  2\,\ln^2 2 + \frac{7\,\zeta_3}{8} \simeq 0.2940 $
\\
$D_{19}$ & $-1$ & $-{3\over 2}-{\pi^2\over 4}+{9\over 2}\ln 2$
 & $-3-\frac{23\,\pi^2}{48} +
  \frac{19\,\ln 2}{2} +
  \frac{\pi^2\,\ln2}{2} -
  \frac{17\,\ln^2 2}{2} -
  \frac{7\,\zeta_3}{2}\simeq -6.0148$ \\
\hline
Total & ${3\over 16}$ & $-{39\over 16}-{\pi^2\over 8}+{3\over 2}\ln 2$
   & $-19.14(27)$
\end{tabular}
\end{center}
\end{table}

\begin{table}
\caption{Comparison of available results for p-Ps and o-Ps.  For the
various coefficients $B$ we use the notation introduced in the text.
$B^{n\gamma}$ describes multiphoton processes, p-Ps $\to 4 \gamma$ and
o-Ps $ \to 5 \gamma$, and $B^{\rm hard}$(LL) is due to effects of
non-linear QED: $\gamma^* \gamma^* \to \gamma \gamma$ for p-Ps and
$\gamma^* \to 3 \gamma$ for o-Ps.  $\Gamma_{\rm LO}$ denotes the
lowest order decay width of a given state.}
\label{tab:comp}
\begin{center}
\begin{tabular}{l r@{.}l l r@{.}l l}

&
\multicolumn{2}{c}{p-Ps}
&
\multicolumn{2}{c}{o-Ps}
\\
\hline
${\cal O}(\alpha)$: coefficients of 
  $\left( {\alpha\over \pi} \right) \Gamma_{\rm LO}$
 & $-2$ & 5325989 & \cite{Harris57} & $-10$ & 286606(10) & \cite{Adkins96}
\\
\hline
${\cal O}(\alpha^2)$:  coefficients of  
  $\left( {\alpha\over \pi} \right)^2 \Gamma_{\rm LO}$
\\

$B^{\rm squared}$ & 1 & 60351 &  & 28 & 860(2) & \cite{Adkins96}
\\
$B^{\rm hard}$(VP) & 0 & 4473430(6) & \cite{AdkShif} & 0 & 964960(4)
& \cite{AdkShif}
\\
$B^{\rm hard}$(LL) & $1$ & 28(13) & [this work] & 0 & 7659(9)
& \cite{AdkShif,AL95}
\\
$B^{\rm hard}(\gamma\gamma)$ & $-\frac {\pi^2}{2\ep}-42$ & 19(27) & [this work]
& \multicolumn{2}{c}{??} &
\\
$B^{\rm soft}$ & $\frac{\pi^2}{2\ep} + 44$ & 002 & [this work]
& \multicolumn{2}{c}{??} &
\\
$B^{\rm total}$     & 5 & 1(3) &  & \multicolumn{2}{c}{??} &
\\
%\hline
$B^{n\gamma}$ & 0 & 274(1) & \cite{AdkBr,Lepage:1983yy} & 0 & 187(11)
& \cite{AdkBr,Lepage:1983yy}
\end{tabular}
\end{center}
\end{table}

\end{document}